\documentclass[a4paper,fleqn]{cas-sc}

\usepackage[authoryear,longnamesfirst]{natbib}
\usepackage{graphicx}
\usepackage{caption}
\usepackage{dcolumn}
\usepackage{bm}
\usepackage{hyperref}
\usepackage[outdir=./]{epstopdf}
\usepackage{amsmath, amsthm}
\usepackage{xcolor}
\usepackage{multirow}
\usepackage{hyperref} 
\usepackage{float}

\usepackage[caption=false]{subfig}
\usepackage{float}
\renewcommand{\printorcid}{}
\def\tsc#1{\csdef{#1}{\textsc{\lowercase{#1}}\xspace}}
\tsc{WGM}
\tsc{QE}
\tsc{EP}
\tsc{PMS}
\tsc{BEC}
\tsc{DE}

\begin{document}
\let\WriteBookmarks\relax
\def\floatpagepagefraction{1}
\def\textpagefraction{.001}
\shorttitle{Applied ML for microstructure evolution}
\shortauthors{A. Pandey and R. Pokharel}

\title [mode=title]{Machine learning enabled surrogate crystal plasticity model for spatially resolved 3D orientation evolution under uniaxial tension}      
\author{Anup Pandey}
\ead{anup@lanl.gov}
\author{Reeju Pokharel}
\ead{reeju@lanl.gov}
\address{Material Science and Technology Division (MST-8), Los Alamos NM, USA}

\begin{abstract}
We present a novel machine learning based surrogate modeling method for predicting spatially resolved 3D microstructure evolution of polycrystalline materials under uniaxial tensile loading. Our approach is orders of magnitude faster than the existing crystal plasticity methods enabling the simulation of large volumes that would be otherwise computationally prohibitive.  This work is a major step beyond existing ML-based modeling results, which have been limited to either 2D structures or only providing average, rather than local, predictions. We demonstrate the speed and accuracy of our surrogate model approach on experimentally measured microstructure from high-energy X-ray diffraction microscopy of a face-centered cubic copper sample, undergoing tensile deformation.

\end{abstract}





\maketitle

\section{Introduction}
Understanding polycrystalline material behaviors requires the observation and study of local hot spots that develop due to complex heterogeneities and non-linearities at the mesoscale. Studying such complex phenomenon based on models requires extremely computationally expensive crystal plasticity (CP) simulations. CP based on finite element analysis frameworks have been developed and extensively used in the past several decades to solve mesoscopic boundary value problems for polycrystalline materials \cite{peirce1983material, anand1994process, bate1999modelling, dawson2000computational, raabe2004using, mayeur2007three, roters2010overview}. While finite element methods (FEM) are more popular, fast Fourier transform (FFT)-based CP methods are also gaining traction due to their lack of meshing requirements and reduced computational cost by solving partial differential equations in Fourier space \cite{lebensohn2001n, lebensohn2012elasto, lebensohn2020spectral}. However, both methods are computationally prohibitive when millions of material simulations are needed to enable advanced material screening and design \cite{franceschetti1999inverse, liu2017materials, butler2018machine}. When extremely large numbers of simulations are required, homogenized methods which have much lower computational cost, can be used instead \cite{lebensohn1993self, miehe1999computational, roters2010overview}. However, homogenized methods only provide average material property predictions, hence local material heterogeneity information due to complex interactions at the grain scale is lost.

Recently, advances in experimental characterization techniques such as high-energy X-ray diffraction microscopy (HEDM) \cite{poulsen2004three, lienert2011high, pokharel2018overview} and Bragg coherent diffractive imaging (BCDI) \cite{gaffney2007imaging, ulvestad2015topological, yau2017bragg} at 3$^{rd}$ and 4$^{th}$ generation light sources have enabled {\it in situ} observation of microstructure and micro-mechanical field evolution of polycrystalline materials in three-dimensions (3D). Such measurements provide unprecedented information on local field evolution under applied stress or temperature. However, these measurements are extremely slow, as a result, material kinetics cannot be studied due to limited temporal resolution. This is mainly due to the large amount of redundant data from multiple projection angles, involving sample rotation, required to perform microstructure reconstruction from measured diffraction patterns \cite{pokharel2018overview}. To improve the data acquisition rates, CP frameworks can be coupled with HEDM data \cite{pokharel2017instantiation}, where CP predictions of microstructure evolution under imposed loading and boundary conditions resembling that of the experiment would eliminate the need for recording redundant data. Such experiments will be able to capture relevant kinetics, while allowing high-fidelity data inversion. Experimental design in conjunction with CP predictions can therefore enable guided beamline experiments, allowing users to make the most out of their limited available beamtime. However, to provide real-time feedback, CP models will have to make instantaneous predictions of microstructure evolution, without requiring several hours or even days of computation in high-performance computing clusters.

The computational cost of these CP models can be lowered through either reduced-order modeling or surrogate modeling. The past decade has seen a significant progress in machine learning (ML) methods that can be utilized for surrogate modeling. Deep learning, convolution neural networks, Gaussian processes, statistical Bayesian inference, and recurrent neural networks have been successfully demonstrated in materials science \cite{farrar2012structural, liu2015predictive, zhang2019extracting}. Data-driven methods have been used to discover new materials with advanced properties \cite{balachandran2016adaptive, mannodi2016machine, yuan2018accelerated}, neural networks have been trained to predict material response to external stimuli \cite{bock2019review}, relate microstructure to mechanical behavior and performance \cite{reimann2019modeling}, and optimize process parameters for additive manufacturing \cite{cook2000combining, baturynska2018optimization}. ML-based algorithms have also been used to solve inverse problems such as reconstructing microstructure from diffraction patterns \cite{cherukara2018real, shen2019convolutional}. 

Mangal et al. \cite{mangal2018applied, mangal2019applied} used random forest learning algorithms to relate stress hot spot development in the grain to its crystallographic properties and local neighborhood. The ML-based framework predicted that a specific grain would develop stress hotspots with $\sim$74$\%$ and $\sim$83$\%$ accuracy for fcc and hcp material systems, respectively, assuming conventional FFT-based CP simulations as the ground truth. Recently, Ali et al. \cite{ali2019application} demonstrated an artificial neural network framework to predict macroscopic material properties such as stress-strain curves and texture evolution in a single crystal, where the ML model showed good agreement with finite element simulations for various loading conditions. There have also been many demonstrations of significant computational gain from employing ML-based surrogate models for solving complex problems instead of using conventional direct numerical simulations approaches \cite{ali2019application, capuano2019smart}. Especially for sequence modeling, recurrent neural networks (RNN) are widely used \cite{mozaffar2018data} but they have shown to be computationally expensive. Recently, a long-short term memory (LSTM) RNN network was proposed \cite{lstm} which can remember the previous states for allowing the modeling of dynamic systems, while being computationally efficient. These networks have been successfully used for material behavior predictions \cite{capuano2019smart, frankel2019prediction}. However, thus far, in attempting to predict material response to imposed loading conditions, the ML frameworks either predict average properties, or only single or a few crystals properties, and only in two-dimensions. 

\subsection{Summary of main results}
In this work, we present a LSTM-based RNN framework for solving a regression problem, to predict the microstructure evolution of polycrystalline materials under tensile stress. We demonstrate for the first time, a state-of-the-art method for generating training sets for developing a surrogate model that provides spatially resolved crystal orientation evolution information in 3D. The main strengths and novelty of this work are:
\begin{enumerate}
    \item The surrogate model is as accurate as the conventional CP model, as well as general enough to predict microstructure evolution of arbitrary 3D representative volume elements (RVE) with previously unseen microstructure parameters such as texture, grain size distribution, and grain morphology. 
    \item The model is trained on synthetic RVEs generated using Dream.3D as input to the FFT-based full-field crystal plasticity simulations, but can be applied to complex experimental data. We demonstrate this capability on a 3D microstructure that was measured using high-energy X-ray diffraction microscopy (HEDM), reported in \cite{pokharel2015situ}.
    \item Once the model was trained, the LSTM-CP approach developed here showed a speed up of $>6\times$ in predicting polycrystalline microstructure evolution under plastic deformation in comparison with EVPFFT for a single 0.02\% strain step and a speed up of $>312\times$ for 1\% strain evolution. 
    \item The model is local in nature, requiring only $3^3$ sized nearest neighborhood data to predict the evolution of any single material point and therefore does not require the use of extremely large and memory-intensive structures as inputs while being size-independent and applicable to any size 3D microstructure volumes.
\end{enumerate}

\section{Methods}

\subsection{Overview of long short-term memory network}
\begin{figure}
	\centering
		\includegraphics[scale=.5]{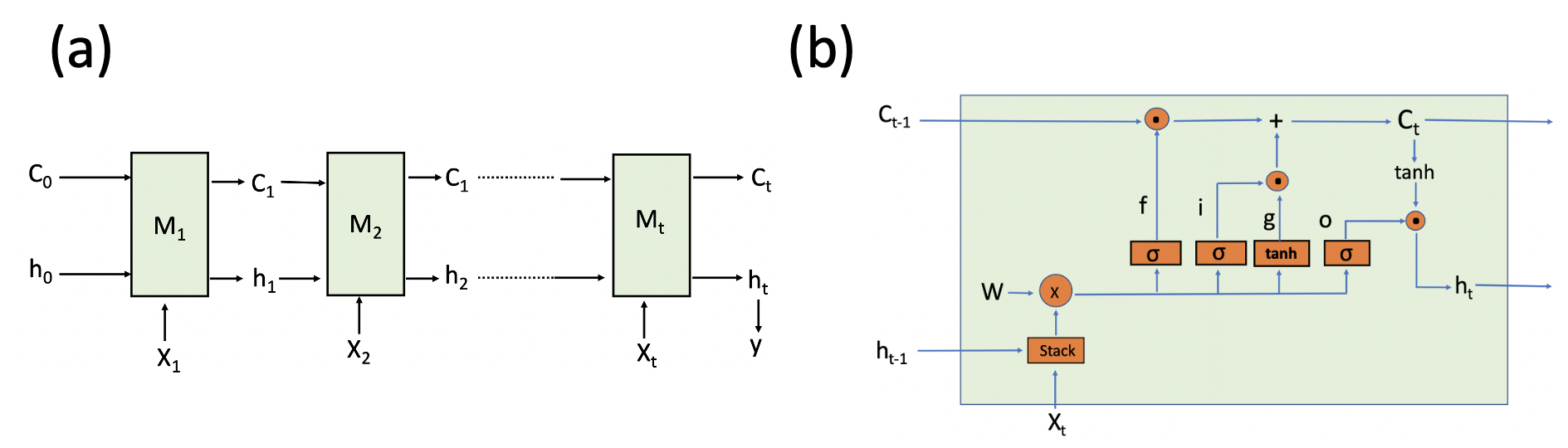} \
	\caption{{\bf (a)} Computational graph for a MISO LSTM network architecture. The $\left \{ X_{1},\dots,X_{t} \right \}$ are a time-series sequence of inputs and $y$ is a single output. The $C_j$ and $h_j$ represent cell states and hidden states, respectively. The $M_j$ are memory blocks. {\bf (b)} Architecture of the LSTM memory block $M_t$ with internal gates $(f_t,i_t,g_t,o_t)$ and matrix weight $W_t$. $C_{t-1}$ and $h_{t-1}$ are the previous cell's state and hidden state and $X_{t}$ is the current input state. $\otimes$ and $\odot$ represent matrix multiplication and element-wise multiplication, respectively.}
	\label{lstm-graph}
\end{figure}

Recurrent Neural Networks (RNN) are a special class of artificial neural networks that contain loops which add feedback and memory to the networks over time. RNNs are, therefore, capable of processing time-series data of dynamic processes evolving over time. RNNs are sometimes susceptible to vanishing and exploding gradients during backpropagation. The gradient issue is addressed by improved algorithms such as  Long short-term memory (LSTM), which is a special type of RNN designed to learn long-term dependencies~\cite{lstm}, with some extra components within a memory cell known as gates. One such multi input single output (MISO) LSTM network architecture is shown in Fig.\ref{lstm-graph}(a), where the inputs are a time-series sequence $\left \{ X_1,\dots,X_t \right \}$ of length $t$, $y$ is a single output, the $\left \{ C_0,\dots,C_t \right \}$ represent cell states, and $\left \{ h_0,\dots,h_t \right \}$ represent hidden states. 

In the MISO LSTM architecture there is one memory block dedicated to each time step $X_j$ where the memory blocks are denoted by $M_j$ with $j\in\left \{ 1,\dots,t\right \}$. The inner loop of the final memory block (all blocks have the same structure) $M_t$ is shown in Fig.\ref{lstm-graph}(b). $M_t$  takes as input one entry of the time-series data, $X_t$ and the previous block's internal cell and hidden states $C_{t-1}$ and $h_{t-1}$. The hidden state $h_{t-1}$ and a current input state $X_{t}$ are stacked together and multiplied by a weight matrix $W_t$ to generate four internal gates $\left ( i_t,f_t,g_t,o_t \right )$, whose functional form is
\begin{eqnarray}
	i_t = f_t = o_t = \sigma \left ( W_t \left [ h_{t-1}, X_{t} \right ]^T \right ), &&   \quad \sigma(x) = \frac{1}{1+e^{-x}}, \nonumber \\
	g_t = \tanh \left ( W_t \left [ h_{t-1}, X_{t} \right ]^T  \right ), && \quad \tanh(x) = \frac{e^x-e^{-x}}{e^x+e^{-x}},
\end{eqnarray}
which are combined with the previous cell state $C_{t-1}$ to generate the current memory block's cell state
\begin{equation}
	C_{t} = f_t\odot C_{t-1}+i_t\odot g_t.
\end{equation}
Finally, the internal cell state $C_{t}$ is used together with the $o_t$ gate to generate the next hidden state
\begin{equation}
	h_{t} = o_t\odot \tanh{C_{t}},
\end{equation}
where $\odot$ represents element-wise matrix multiplication. 
For our application the $\left \{ X_1,\dots,X_t \right \}$ represent snapshots of the orientations of a 3D material volume undergoing evolution from one strain state to the next. Our goal is to use this sequence of data to predict the orientation of the material at the next state $y_{t+1}$, which is represented by the network's output $y$. 

\subsection{Data preparation for training LSTM network}
Training neural networks for developing surrogate models requires very large data sets of high resolution measurements. It is prohibitively time consuming to obtain large numbers of high temporal and spatial resolution sequences of 3D microstructure evolution data directly from limited beam time experiments at over subscribed light sources such as the APS. We have attempted to mitigate this need for a large amount of experimental data by generating synthetic data to be used as inputs for training the LSTM model. The data set required for training the LSTM network was prepared in three steps as described below.

\subsubsection{3D Microstructure generation and evolution}
\begin{figure}
	\centering
		\includegraphics[scale=.35]{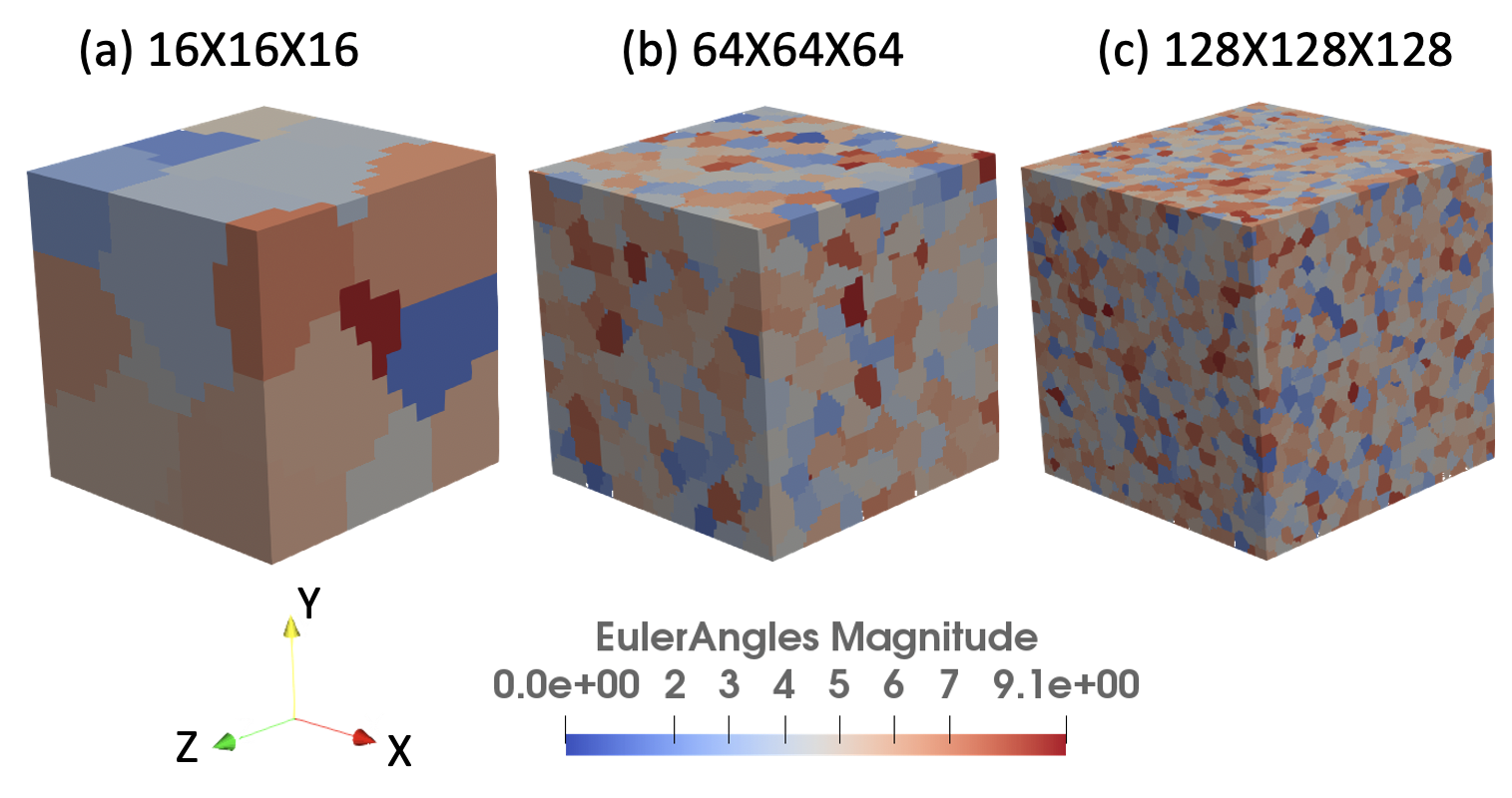} \
	\caption{Representative synthetic 3D microstructures discretized on (a) 16$^3$, (b) 64$^3$, and (c) 128$^3$ grids are generated using Dream.3D, where crystal orientation (defined by an ordered set of three Euler angles) are assigned to each voxel. Colors represent the Euler angles magnitude.}
	\label{syn-str}
\end{figure}

In the first step of data preparation, synthetic 3D microstructures were generated using Dream.3D software~\cite{dream3d}. We generated 100, 20, and 5 instances of representative 3D microstructures discretized on $16\times16\times16$, $64\times64\times64$, and $128\times128\times128$ grids, with average numbers of grains of 25, 1315 and 10071, respectively. An image of each representative structure is shown in Fig. \ref{syn-str}. In the second step, microstructure evolution data were generated from running a numerical simulation of deformation under uniaxial tensile loading, by instantiating the full field elasto-viscoplastic fast Fourier transform (EVPFFT) crystal plasticity model developed by ~\cite{ricardo} with a synthetic 3D microstructure. The details of the EVPFFT model and model instantiation with 3D microstructure for numerical simulations are given elsewhere ~\cite{ricardo, pokharel2017instantiation}. The model parameters such as elastic stiffness constants and Voce hardening parameters used for numerical simulations were
\begin{eqnarray}
	&& C_{11} \ [\mathrm{GPa}]  = 168.4, \quad C_{12} \ [\mathrm{GPa}] = 121.4, \quad C_{44} \ [\mathrm{GPa}] = 75.4, \nonumber \\
	&& \tau_0 \ [\mathrm{MPa}] = 45.0, \quad \tau_1 \ [\mathrm{MPa}] = 30.0, \quad \theta_0 = 800.0, \quad \theta_1 = 130.0. \nonumber
\end{eqnarray}
Each simulation was carried out for up to 12\% strain at a rate of 0.02\% strain on each step.  
%

\subsubsection{Constructing sequence data}
\begin{figure}
	\centering
		\includegraphics[scale=.35]{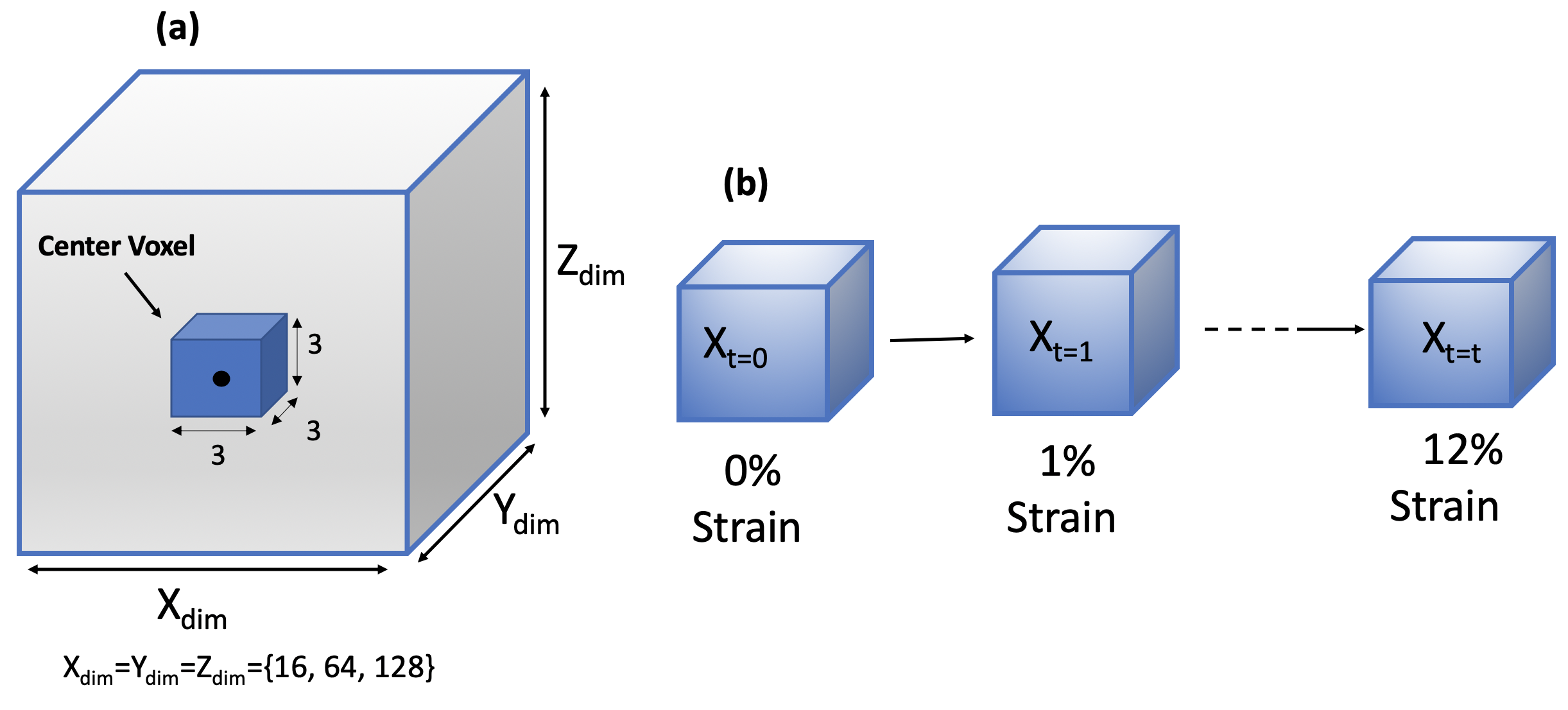}
	\caption{({\bf a}) Neighboring $3\times3\times3$ voxels for the center voxel for a representative microstructure. ({\bf b}) Evolution of the center voxel and its neighbors on the application of strains.}
	\label{microstr_seq}
\end{figure}

To accurately predict spatially resolved 3D microstructure evolution, it was important to formulate the problem in such a way that the method learned the local physics, and at the same time did not have to rely on using large volumes as inputs during training. Towards developing such a data-driven surrogate CP solver, we implemented a novel approach that considered only the effect of the local neighborhood on each material point (voxel). We prepared a sequence of data for individual voxels consisting of $3\times3\times3$ cubic neighborhoods of the microstructure in which they were centered, as shown in Fig.\ref{microstr_seq} (a). For predicting the evolution of a single voxel, we used only the information from a $3^3$ cube at which the voxel was centered (up to the third nearest neighbor). We did not make predictions for voxels on the edges of the RVE so that the local neighborhood around each voxel was always fully contained within the volume. Therefore each $N^3$ RVE contained $(N-2)^3$ useful $3^3$-sized training volumes. For instance, a single $128^3$ structure provided $\sim2\times10^6$ training locations. This approach drastically increased the number of data points for model training while significantly reducing the number of expensive crystal plasticity simulations of 3D RVEs. This approach also eliminated the use of large volumes as inputs during training. 

Next, we prepared a sequence of 13 strain steps (including the original undeformed state) at an interval of 1\% strain increment; therefore, each voxel in the microstructure had 13 steps in its sequence (Fig.\ref{microstr_seq} (b)). For each voxel V$_{t}^{j}$ at a given position $j$ and time step $t$, a $3\times3\times3$ subset of its neighboring voxels X$_{t}^{j}$ was determined. Therefore, in a given sequence, V$_0^{j}$,V$_1^{j}$,.....,V$_{12}^{j}$ are the center voxel and X$_0^{j}$, X$_1^{j}$,....,X$_{12}^{j}$ are the corresponding neighboring voxels at different levels going from 0\% to 12\% tensile strains. The data are arranged in ({\bf X},{\bf Y}) format as shown in Table \ref{tbl3}, where {\bf X} is the neighborhood of voxels around a central voxel at a given strain level $t$ and {\bf Y} is the central voxel strained by 1\% incremental strain at a given strain level $(t+1)$. A crystal orientation represented by three Euler angles ($\phi_1$,$ \Phi$, $\phi_2$) was assigned to each voxel. Therefore, each {\bf X} is an 81 component ($3\times3\times3\times3$) vector and each {\bf Y} is an output vector with 3 Euler angles. As mentioned earlier, for simplicity, we have ignored the edge voxels in the training and test sets used in our model and have considered only those voxels with 26 nearest neighbors surrounding it.
%
\begin{table}[width=.5\linewidth,cols=2,pos=h]
\caption{Training data format for a series of time steps for each voxel at a given position $j$ in the RVE.}\label{tbl3}
\begin{tabular*}{\tblwidth}{@{} LL@{} }
\toprule
{\bf X} & {\bf Y} \\
\midrule
 X$_0^{j}$ & V$_1^{j}$ \\
 X$_1^{j}$ & V$_2^{j}$ \\
 X$_2^{j}$ & V$_3^{j}$ \\
$\vdots$& $\vdots$\\
 X$_{11}^{j}$ & V$_{12}^{j}$ \\
\bottomrule
\end{tabular*}
\end{table}

%

\subsection{LSTM models}
Arrangement of data, as shown in Table \ref{tbl3}, is in a univariate series, where LSTM learned to predict the output observation based on the previous observation of the series. The training data sets consisted of multiple one-step ({\bf X},{\bf Y}) sequences in strain rate, up to 12\% strain. We trained vanilla LSTM models, with a single LSTM cell and 50 hidden layers, the most straightforward form of LSTM architecture.  

We took our LSTM architecture and developed two predictive models, referred to as Model I and Model II, by training the weights on two different data sets. Model I was trained using the 16$\times$16$\times$16 synthetic microstructures and their EVPFFT simulated structures up to 12\% strain, at a strain increment of 1\%. Model II was trained using 16$\times$16$\times$16, 64$\times$64$\times$64 and 128$\times$128$\times$128 mixed synthetic structures and their EVPFFT simulated structures up to 12\% strain, at a strain increment of 1\%. To study the effect of strain increment size in the accuracy of the model prediction, we also trained two more models with a strain step size of 2\% (Model III) and 3\% strain (Model IV) in the sequential training data consisting of only 16$\times$16$\times$16 synthetic microstructures. A summary of model data and strain step sizes is given in Table \ref{tbl2}.

\begin{table}[width=.9\linewidth,cols=4,pos=h]
\caption{Model Characteristics}\label{tbl2}
\begin{tabular*}{\tblwidth}{@{} LLLLL@{} }
\toprule
Model & Model $\mathrm{I}$ & Model $\mathrm{II}$ & Model $\mathrm{III}$ & Model $\mathrm{IV}$ \\
\midrule
Training Data Volume & $16^3$ & $16^3$, $64^3$, $128^3$ & $16^3$ & $16^3$ \\
\midrule
Strain Step Size & 1\% & 1\% & 2\% & 3\% \\

\bottomrule
\end{tabular*}
\end{table}

EVPFFT simulated microstructures at strain levels up to 12\% were considered as the ground truth. The four models are referred to as LSTM-CP from hereinafter. For Model I, Model III and Model IV, there were 658560 and 164640 number of data points in the training and test sets, respectively. For Model II there are 111782780 and 7452980 number of data points in the training set and the test set, respectively. We used logcosh as a loss function between each Euler angle component of the model predictions and the ground truth. The Adam~\cite{kingma2014adam} optimizer with a learning rate of 0.00001 was used in training the models. We used disorientation angle as a final metric for comparing LSTM-CP predictions with the ground truth. Disorientation angle is the smallest possible rotation angle out of all symmetrically equivalent misorientations that lie within a fundamental zone (FZ), and is given by:
\begin{equation*}
    \Delta g_{AB}= O_{B} g_{B}(O_{A} g_{A})^{-1},
\end{equation*}
where O denotes one of the cubic symmetry operators, and g$_{A}$ and  g$_{B}$ are the orientations from the `ground truth' and LSTM-CP model predictions, respectively. 

\section{Results}
\begin{figure}
	\centering
		\includegraphics[scale=.70]{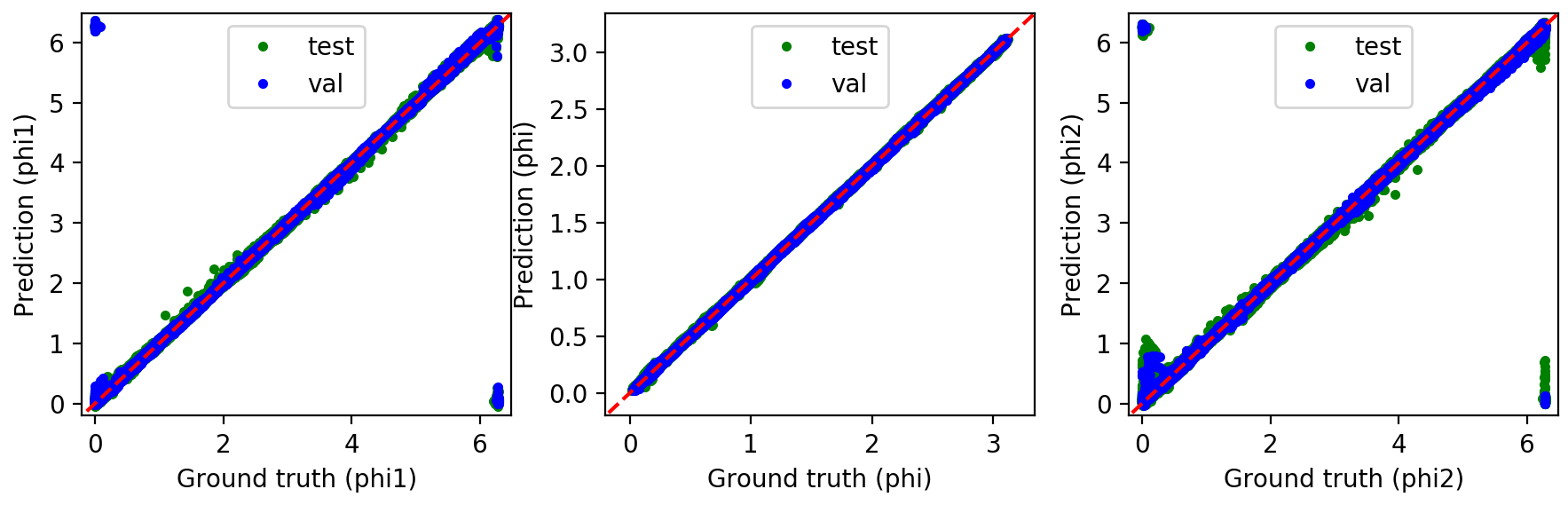}
	\caption{The LSTM-CP predicted Euler angles versus the ground truth (FFT) for training dataset from Model I. Model I is trained from the 16$\times$16$\times$16 structures.}
	\label{lstmpred_m1}
\end{figure}

\begin{figure}
	\centering
		\includegraphics[scale=.70]{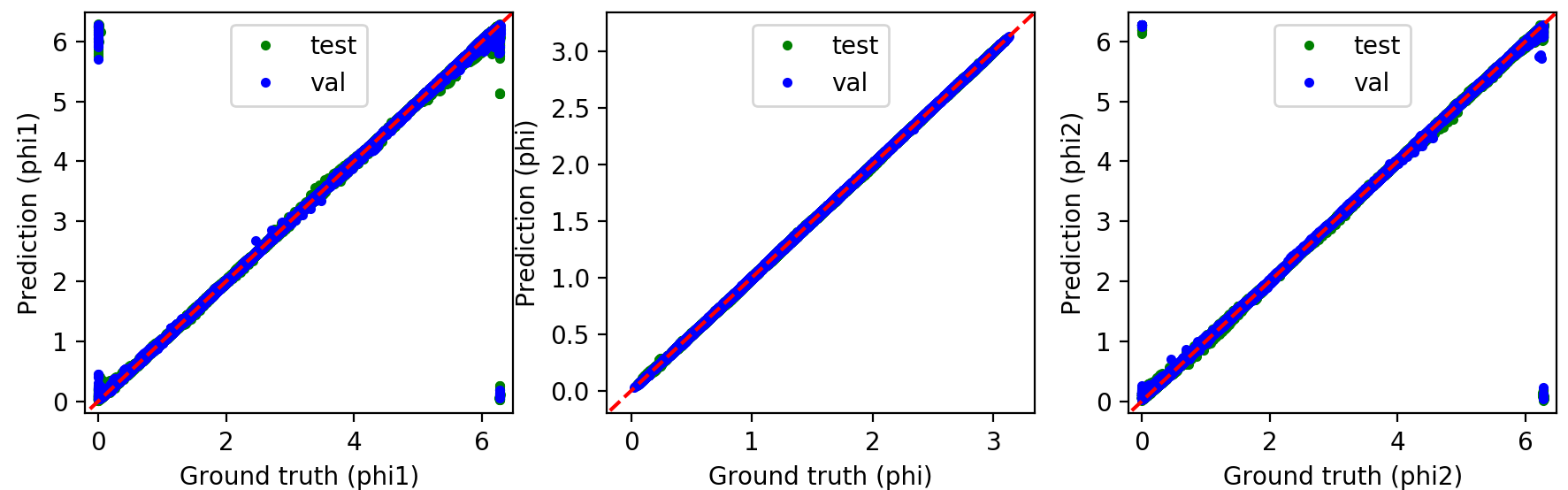}
	\caption{The LSTM-CP predicted Euler angles versus the ground truth (FFT) for training dataset from Model II. Model II is trained from  the 16$\times$16$\times$16, 64$\times$64$\times$64 and 128$\times$128$\times$128  mixed structures.}
	\label{lstmpred_m2}
\end{figure}

Model I and Model II predictions for the test set and the validation set at the end of the training process are shown in Fig. \ref{lstmpred_m1} and Fig. \ref{lstmpred_m2}, respectively. The model prediction of three orientation angles ($\phi_1$, $\Phi$, $\phi_2$) is excellent, with only a few outliers. 

%

\subsection{LSTM-CP models prediction for the 3D synthetic structures}
We predicted full 3D microstructures evolution (excluding the edges and not used in training) by 1\% strain using the LSTM-CP models for 16$^3$ and 64$^3$ synthetic structures. The fidelity of LSTM-CP model predictions of the voxel orientations after deformation was measured by the disorientation angle.

\subsubsection{16$\times$16$\times$16 3D synthetic structure}
\begin{figure}
\centering
\subfloat[Disorientation distribution between the LSTM-CP Model I prediction and the EVPFFT simulations of synthetic structure from no strain (0\%) to 1\% strain. The mean is 1.31$^{\circ}$ and the standard deviation of the distribution is 0.821$^{\circ}$. \label{misorsyn}]{%
\includegraphics[scale=.50]{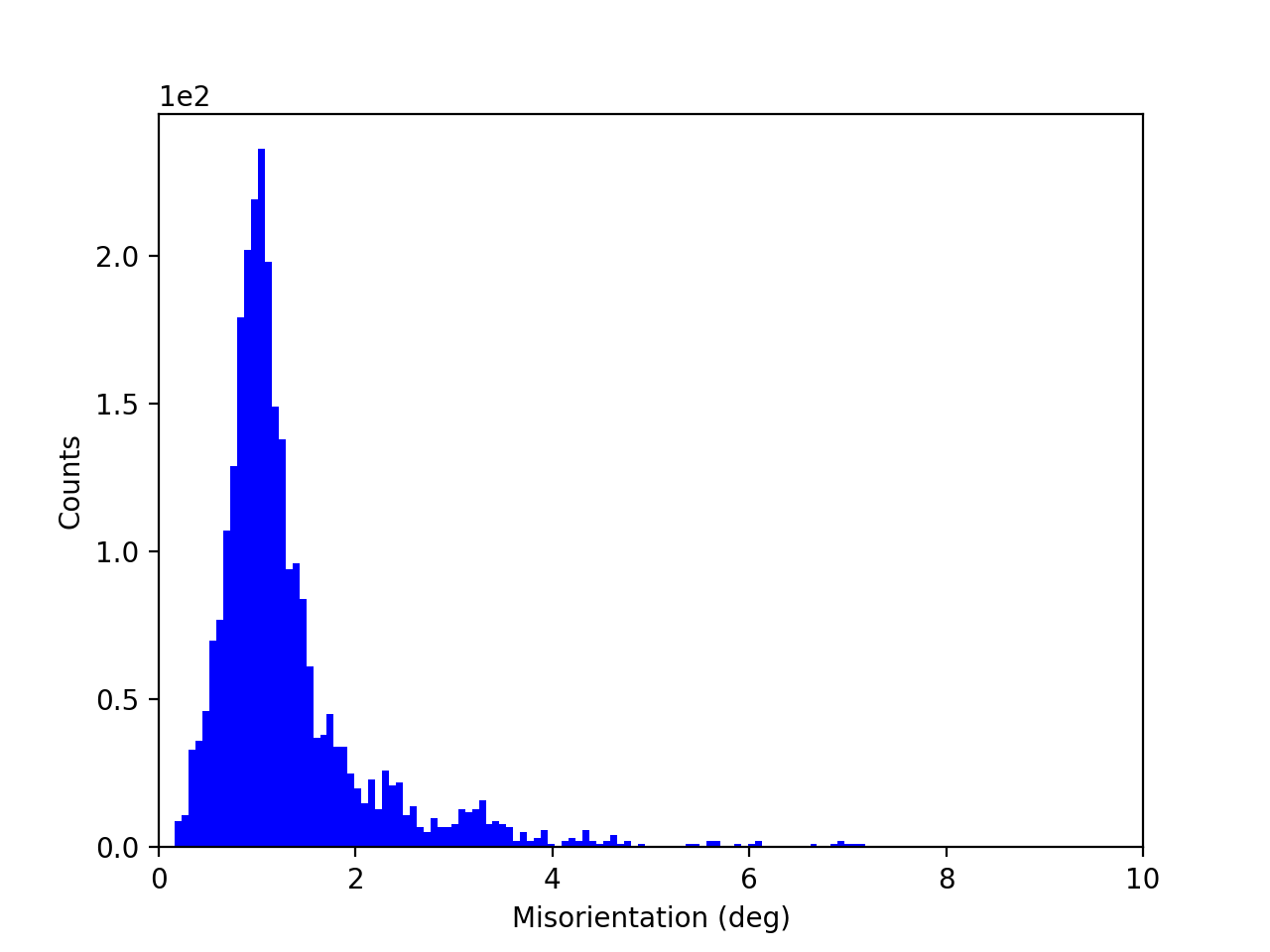}
 }\qquad
 \subfloat[Disorientation distribution between the LSTM-CP Model II prediction and the EVPFFT simulations of synthetic structure from no strain (0\%) to 1\% strain. The mean of the distribution is 0.96$^{\circ}$ and standard deviation is 0.601$^{\circ}$. \label{misorsyn_mixed}]{%
\includegraphics[scale=.50]{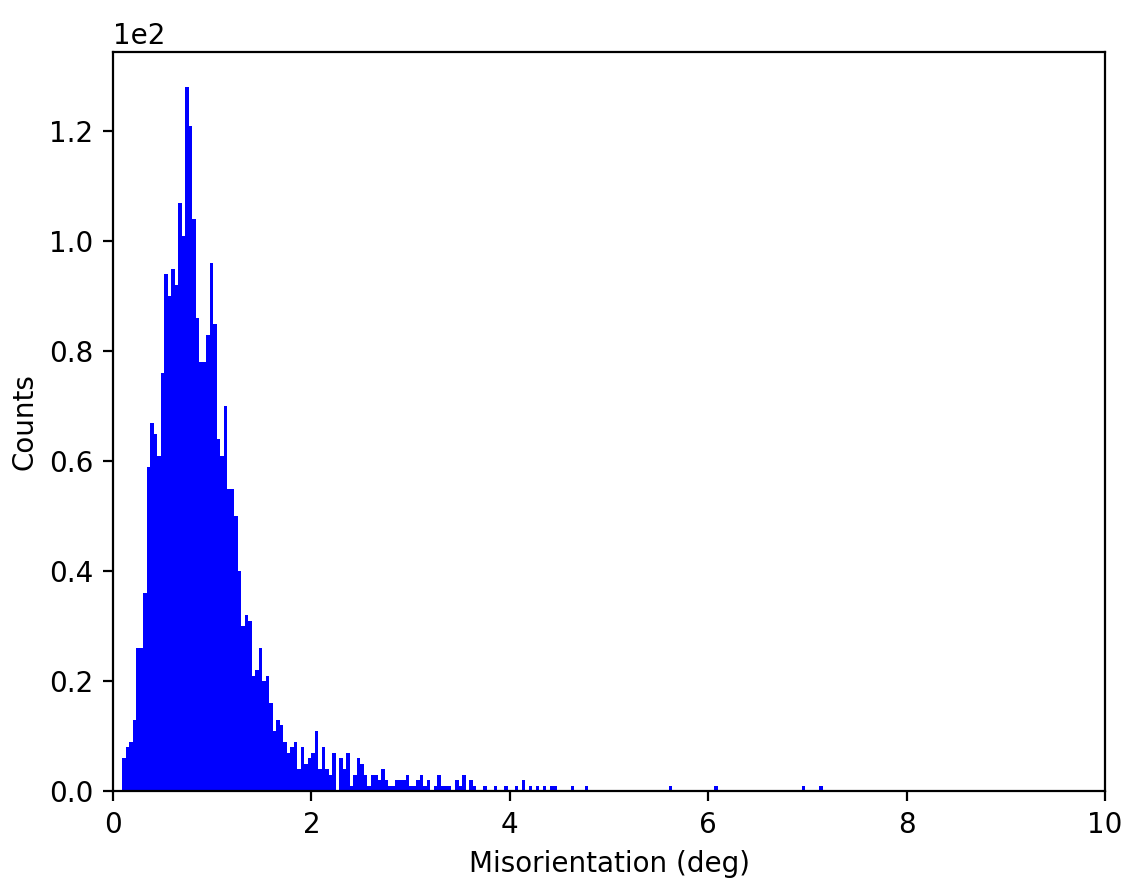}
 }\qquad
 \subfloat[ The disorientation map between the ground truth and LSTM-CP Model I predicted from 0\% to 1\% strain for the different layers (layer 4, 8 and 12) of the 3D microstructure structure  shown in Fig.\ref{syn-str}.\label{misormap}]{%
\includegraphics[scale=.60]{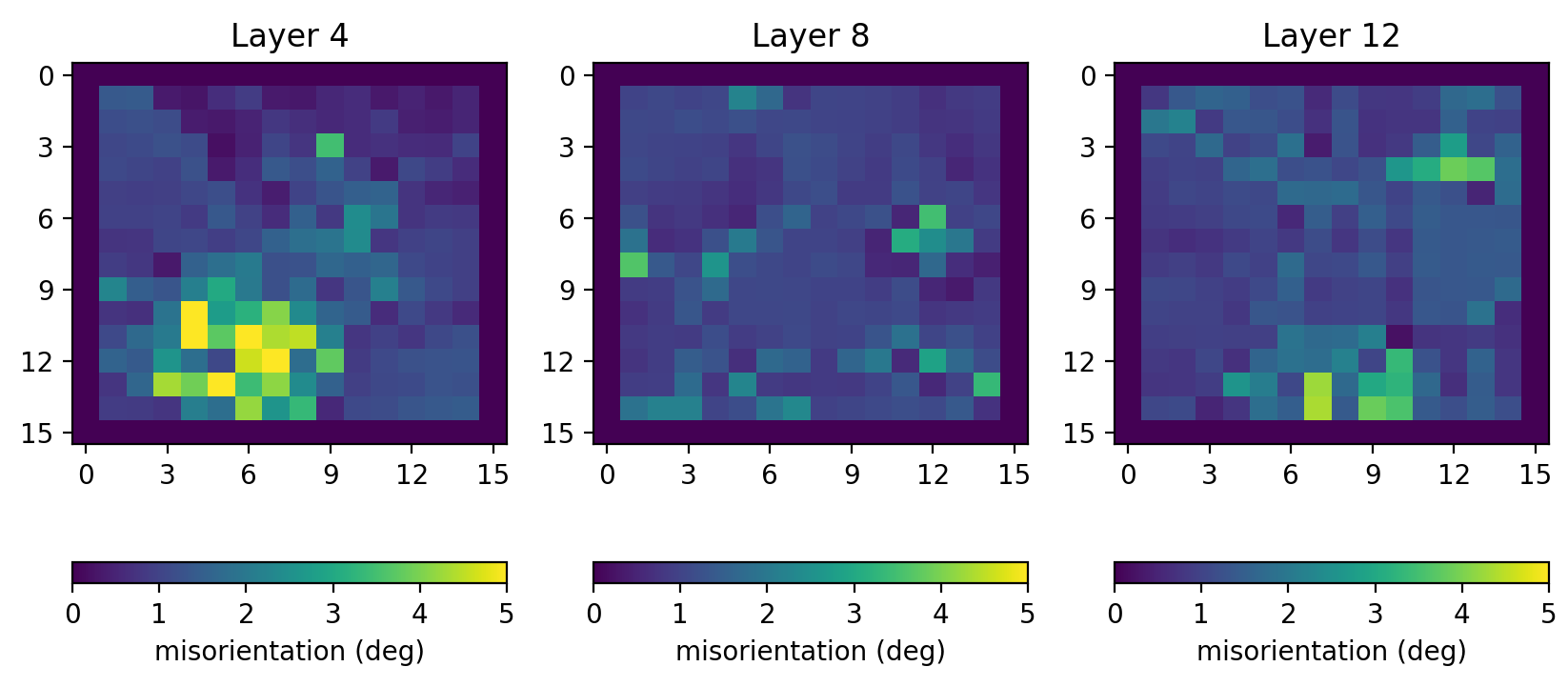}
 }\qquad
\subfloat[The disorientation map between the ground truth and LSTM-CP Model II predicted from 0\% to 1\% strain for the different cross sections (layer 4, 8 and 12) of the 3D microstructure structure  shown in Fig.\ref{syn-str}.\label{misormap_mixed}]{%
\includegraphics[scale=.60]{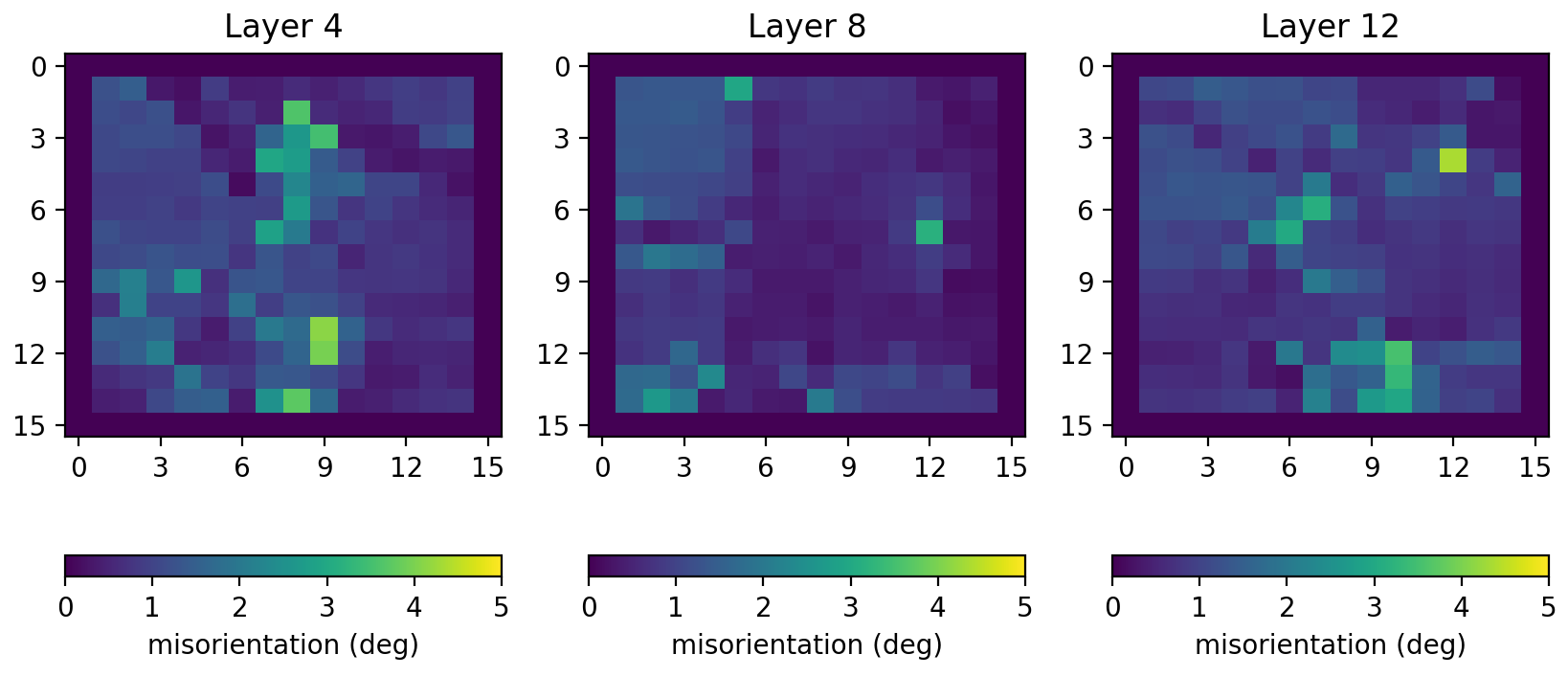}
  }%
\caption{Disorientation distribution for the LSTM-CP Model I and Model II predicted microstructure and the misorientation maps for different cross sections (layers 4, 8 and 12) for the synthetic microstructure.}
\label{misor_16cube}
\end{figure}

For 16$^3$ 3D synthetic structure, the disorientation angle between the ground truth and the LSTM-CP Model I and Model II predicted orientations are calculated, and the distributions are shown in Fig.\ref{misorsyn} and Fig.\ref{misorsyn_mixed}, respectively. For Model I and Model II, predictions from no strain (0\%) to 1\% strain, 99.38\% and 99.85\% are within 5$^{\circ}$ disorientation angles, respectively. For Model I, the mean of the distribution is 1.31$^{\circ}$ and the standard deviation is 0.821$^{\circ}$, and for Model II, the mean is 0.96$^{\circ}$ and standard deviation is 0.601$^{\circ}$. The smaller mean disorientation angle implies high fidelity prediction. The spatially resolved misorientation map between the ground truth and LSTM-CP Model I and Model II predicted from 0\% to 1\% strain for the different layers (layers 4, 8, and 12) of the 16$^3$ 3D structure are shown in Fig.\ref{misormap} and Fig.\ref{misormap_mixed}, respectively. For both the models, all three layers corroborate the overall low disorientation angle distribution. The comparison of misorientation maps shows that Model II is slightly better than Model I. The results elucidate the robustness of LSTM-CP models in precisely capturing the spatial distribution of orientation evolution in the synthetic structure to the level of crystal plasticity simulations.  
 
\begin{figure}
\centering
\subfloat[LSTM-CP Model I \label{stat_syn161616}]{%
\includegraphics[scale=.70]{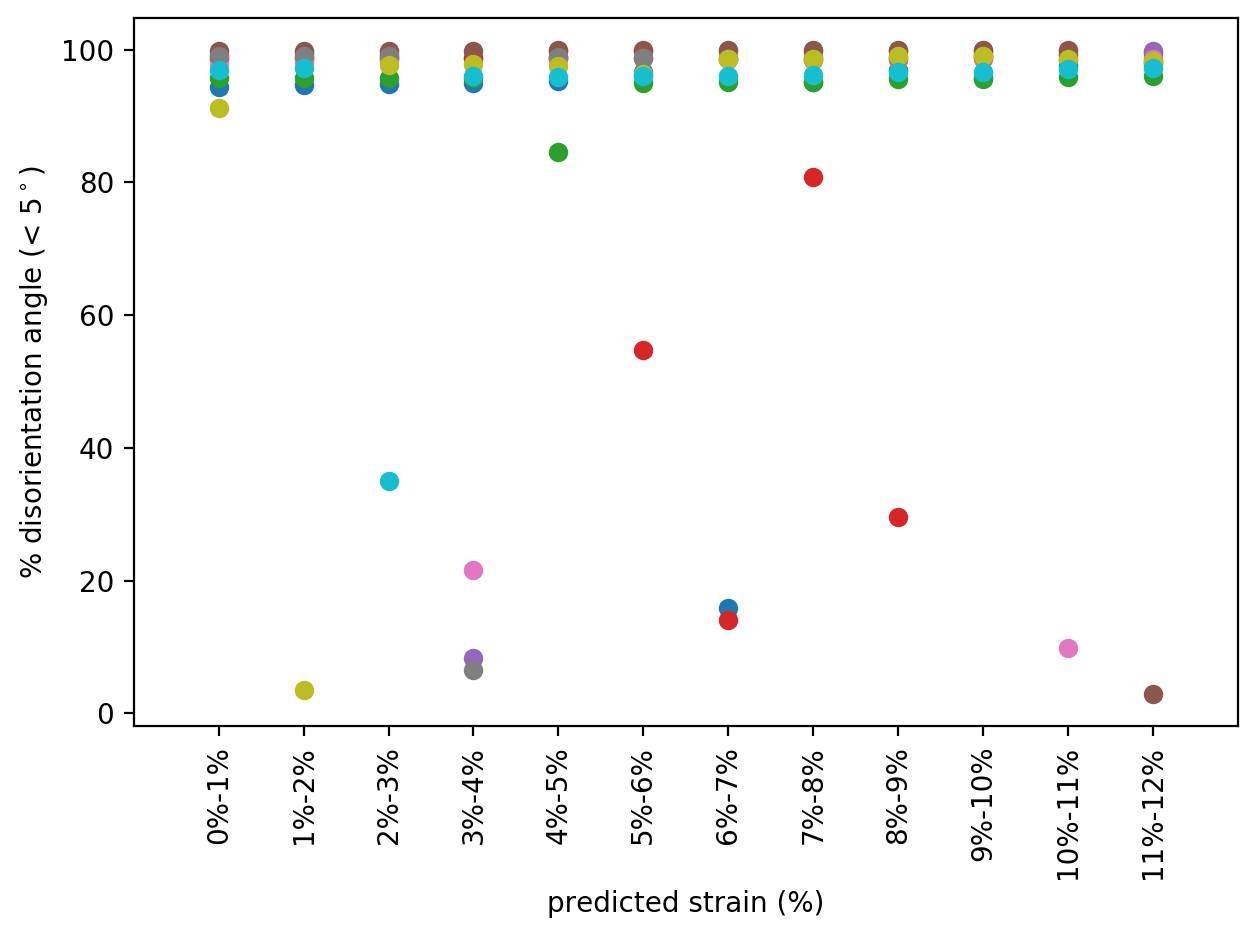}
 }\qquad
\subfloat[LSTM-CP Model II \label{stat_syn161616_mixed}]{%
\includegraphics[scale=.70]{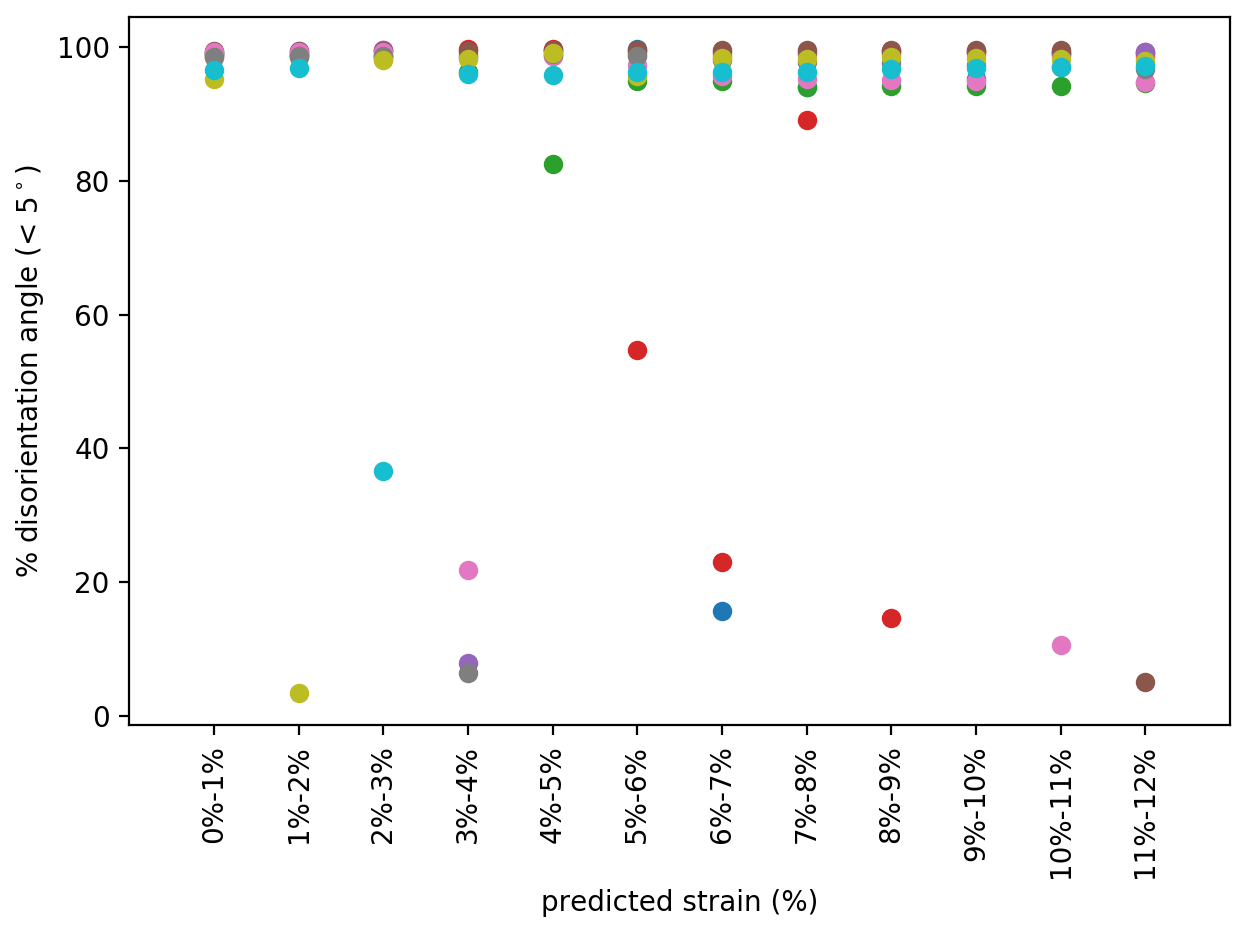}
  }%
\caption{Percentage of disorientation angles less than 5$^{\circ}$ for 10 different 16$\times$16$\times$16 synthetic microstructures for the LSTM-CP Model I  and Model II predictions at different strain state. The numbers in X-axis corresponds to the evolution from the preceding strain state}
\label{stat_syn161616_all}
\end{figure}

The models can predict the evolution of a microstructure by 1\% strain from any given strain state up to 12\% strain. We predicted the evolution of 10 random microstructures up to the 12\% strain for both the models and the percentage of disorientation within 5$^{\circ}$ is shown in Fig.\ref{stat_syn161616_all}. The initial state is taken as the EVPFFT simulated structure for which the evolution of 1\% strain is predicted by the models (e.g., for 5\% EVPFFT simulated initial structure, the models predict the 6\% strained structure and are compared with the 6\% EVPFFT simulated structure or ground truth). Both models are following a similar prediction trend in the sense that predictions for some of the strain levels for some structures have lower accuracy with large disorientation angles. For most of the structures, the predictions from both the models show $>$98\% voxels within 5$^{\circ}$ disorientation angles (Fig.\ref{stat_syn161616} and Fig.\ref{stat_syn161616_mixed}).

%

%

\subsubsection{64$\times$64$\times$64 3D synthetic structure}
\begin{figure}
\centering
\subfloat[Disorientation between the LSTM-CP Model I prediction and the EVPFFT simulations of 64$\times$64$\times$64 synthetic structure from no strain (0\%) to 1\%. 99.43 \% of disorientations are within 5$^\circ$. The mean is 1.04$^{\circ}$ and the standard deviation of the distribution is 0.916$^{\circ}$. \label{misorsyn_64cube}]{%
\includegraphics[scale=.45]{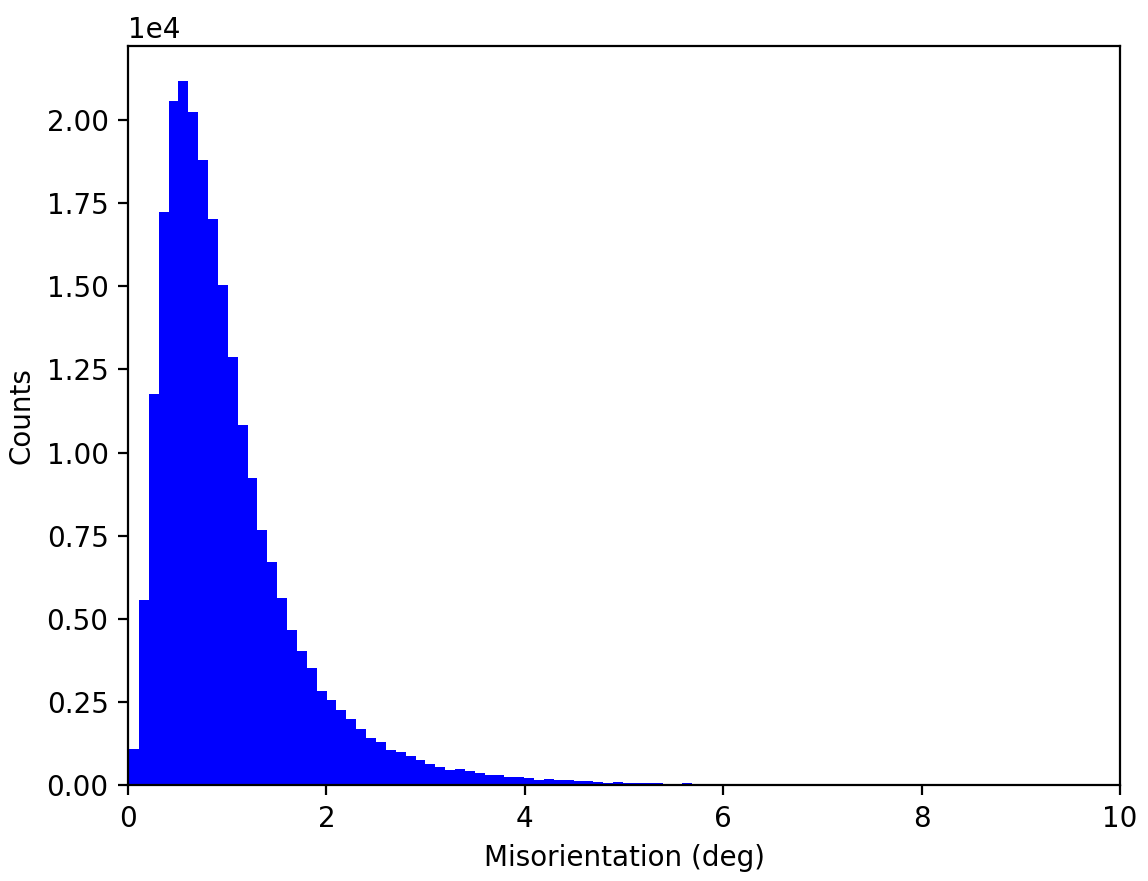}
 }\qquad
 \subfloat[Disorientation between the LSTM-CP Model II prediction and the EVPFFT simulations of 64$\times$64$\times$64 synthetic structure from no strain (0\%) to 1\%. 99.29 \% of disorientations are within 5$^\circ$. The mean is 1.12$^{\circ}$ and the standard devaition of the distribution is 0.852$^{\circ}$. \label{misorsyn_64cube_mixed}]{%
\includegraphics[scale=.45]{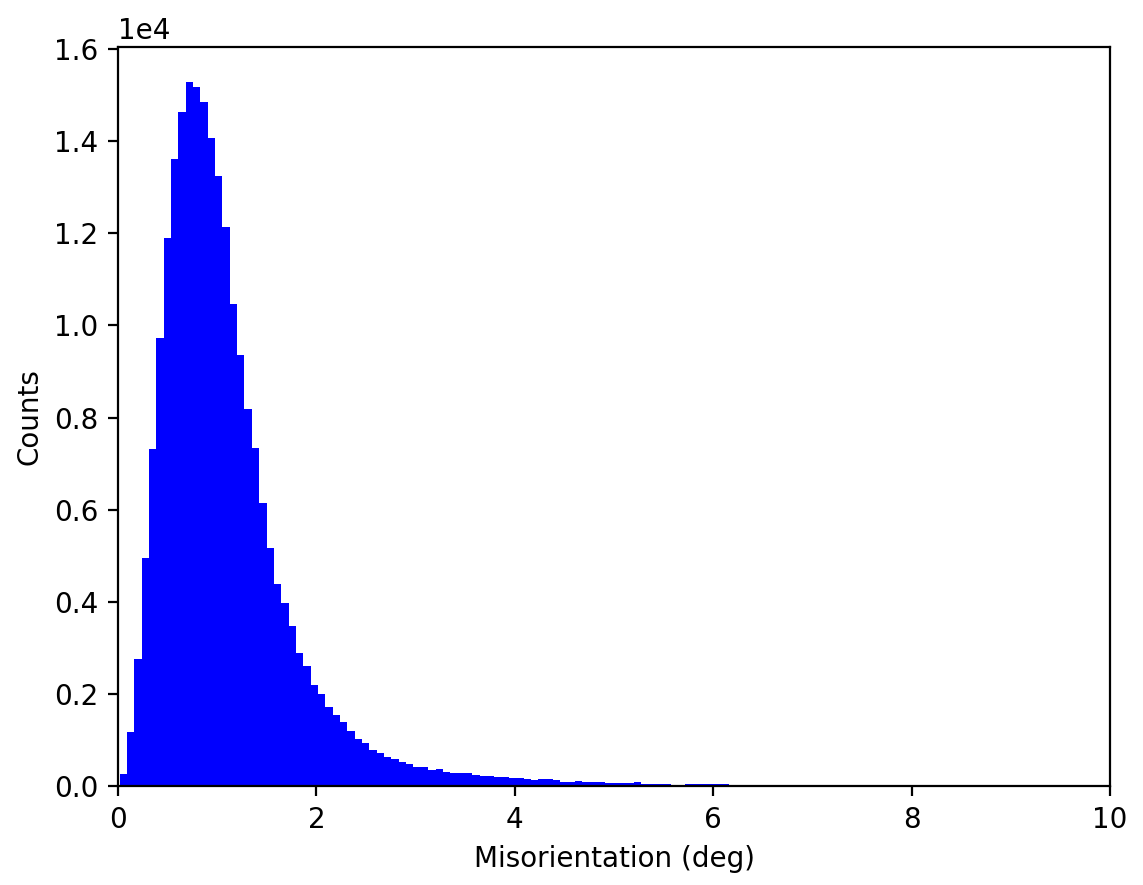}
 }\qquad
 \subfloat[The misorientation map between the ground truth and LSTM-CP Model I predicted from 0\% to 1\% strain for the different layers (layer 4, 8 and 12) of the 64$\times$64$\times$64 3D microstructure  \label{misormap_64cube}]{%
\includegraphics[scale=.65]{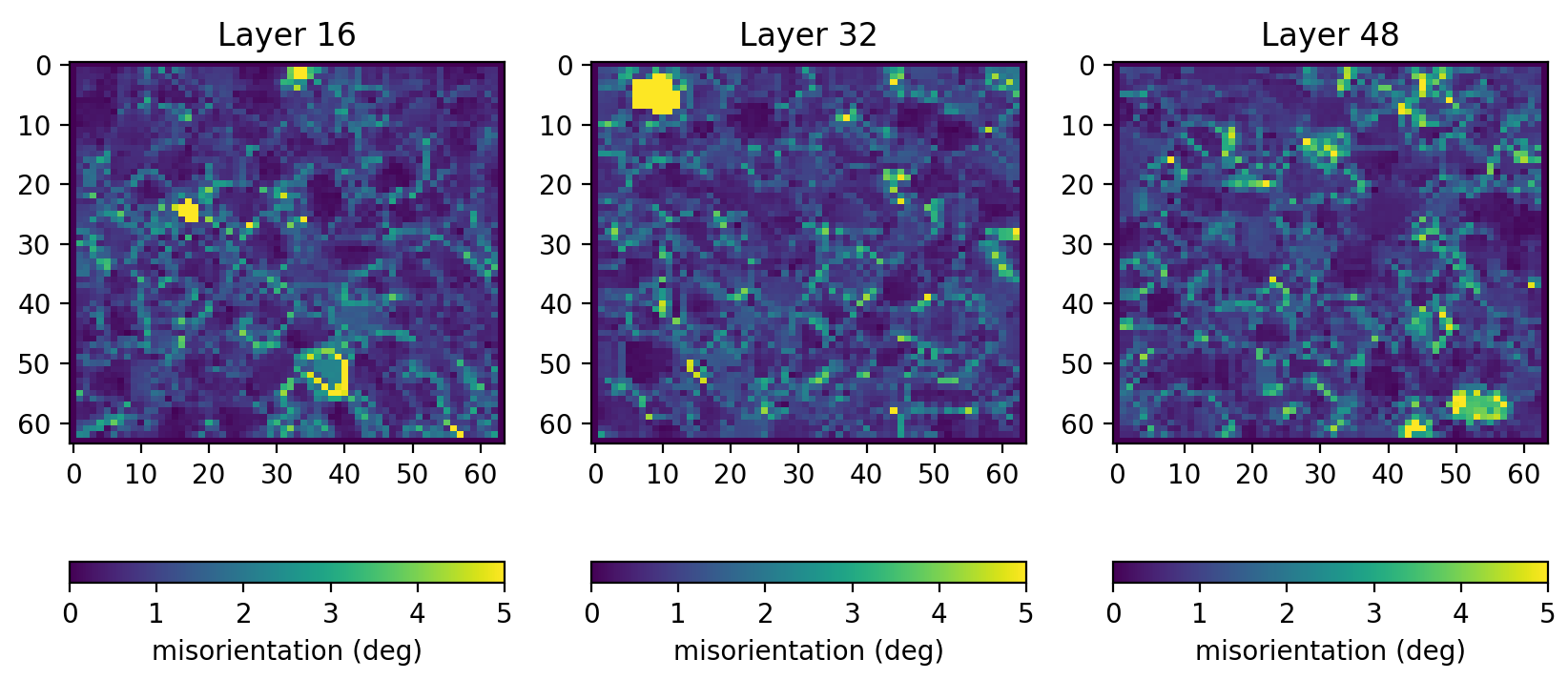}
 }\qquad
\subfloat[The misorientation map between the ground truth and LSTM-CP Model II predicted from 0\% to 1\% strain for the different layers (layer 4, 8 and 12) of the 64$\times$64$\times$64 3D microstructure .\label{misormap_64cube_mixed}]{%
\includegraphics[scale=.65]{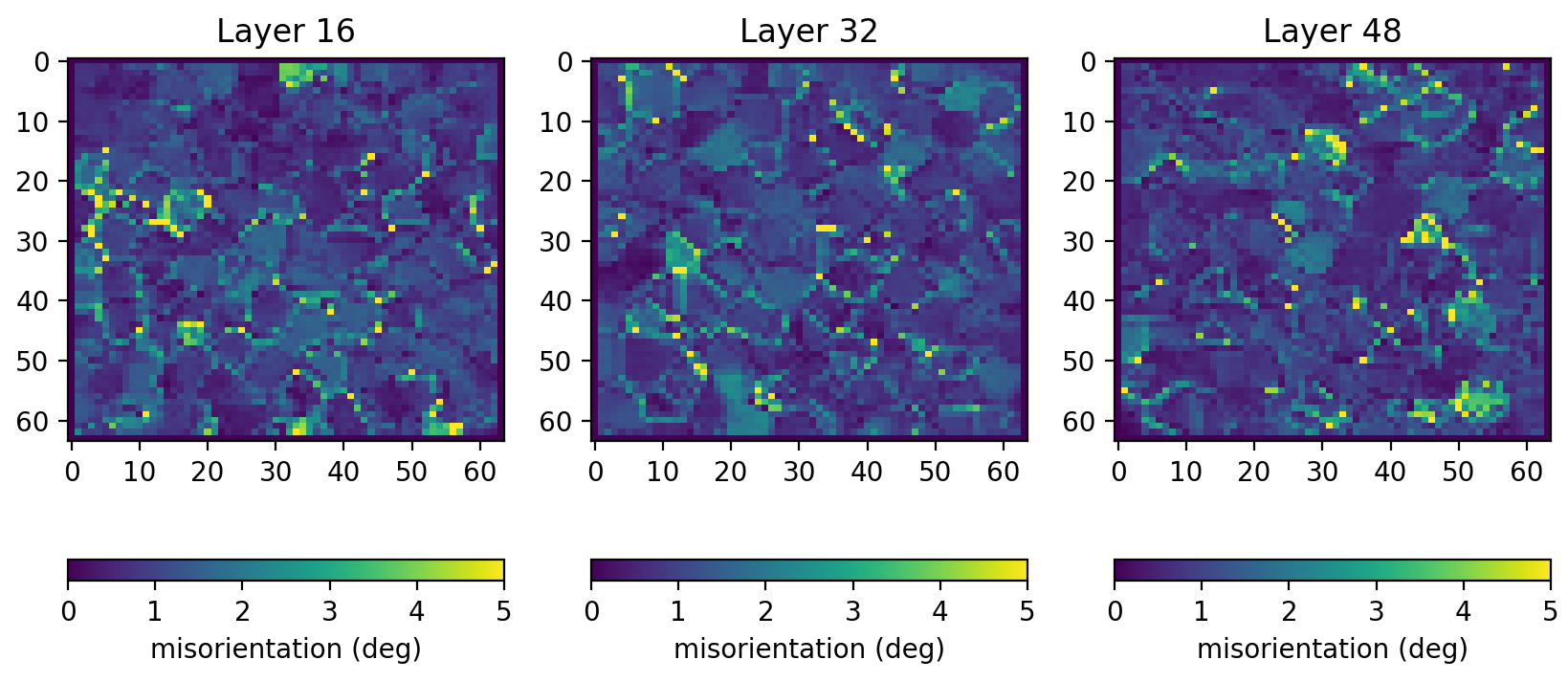}
  }%
\caption{Disorientation distribution for the LSTM-CP Model I predicted microstructure and the misorientation maps for layers 4, 8 and 12 for a random 64$\times$64$\times$64 synthetic microstructure.}
\label{misor_64cube}
\end{figure}

\begin{figure}
	\centering
		\includegraphics[scale=.50]{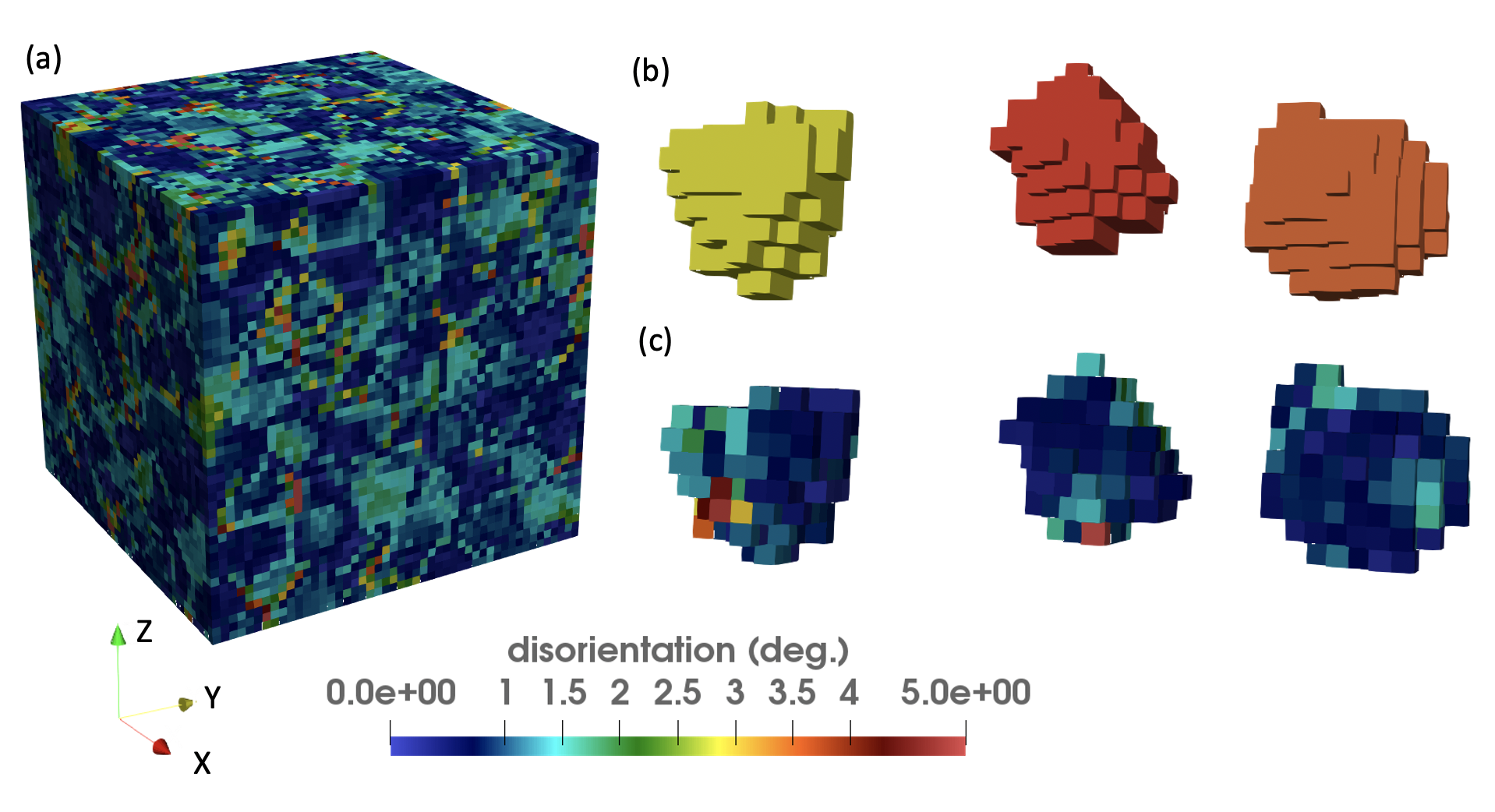}
	\caption{(a) Distribution of disorientation between 1\% EVPFFT simulated 64$\times$64$\times$64 synthetic microstructure  and the LSTM-CP Model II. Color scheme shows the disorientation angle in degrees. (b) Orientations of random grains. (c) Disorientation distribution in random grains shown in (b).}
	\label{grain_64cube}
\end{figure}

 The orientations evolution of full 3D 64$^3$ synthetic microstructure from no strain (0\%) to 1\% strain are predicted using both the LSTM-CP Model I and Model II. The disorientation angle between the orientations from the ground truth (EVPFFT simulated to 1\% strain) and model predictions are calculated, and the distributions are shown in Fig.\ref{misorsyn_64cube} and Fig.\ref{misorsyn_64cube_mixed}, respectively. For Model I and Model II, predictions from no strain (0\%) to 1\% strain, 99.43\% and 99.29\% of the disorientations are within 5$^{\circ}$. For Model I, the mean of the distribution is 1.04$^{\circ}$ and the standard deviation is 0.916$^{\circ}$, and for Model II, the mean is 1.12$^{\circ}$ and the standard deviation is 0.852$^{\circ}$. The corresponding 2D maps of three different cross sections (layers 16, 32 and 48) of a synthetic microstructure are shown in Fig.\ref{misormap_64cube} and Fig.\ref{misormap_64cube_mixed}, respectively. The 2D maps corroborate narrow disorientation angle distributions. Both the model predictions are qualitatively and quantitatively similar. We also plotted the disorientation between the LSTM-CP Model II predictions and the ground truth for three random 3D grains extracted from the 64$^3$ structure, which is shown in Fig.\ref{grain_64cube}. The grain scale map also confirms the small disorientation angles. 
 
\begin{figure}
\centering
\subfloat[LSTM-CP Model I \label{stat_syn646464}]{%
\includegraphics[scale=.65]{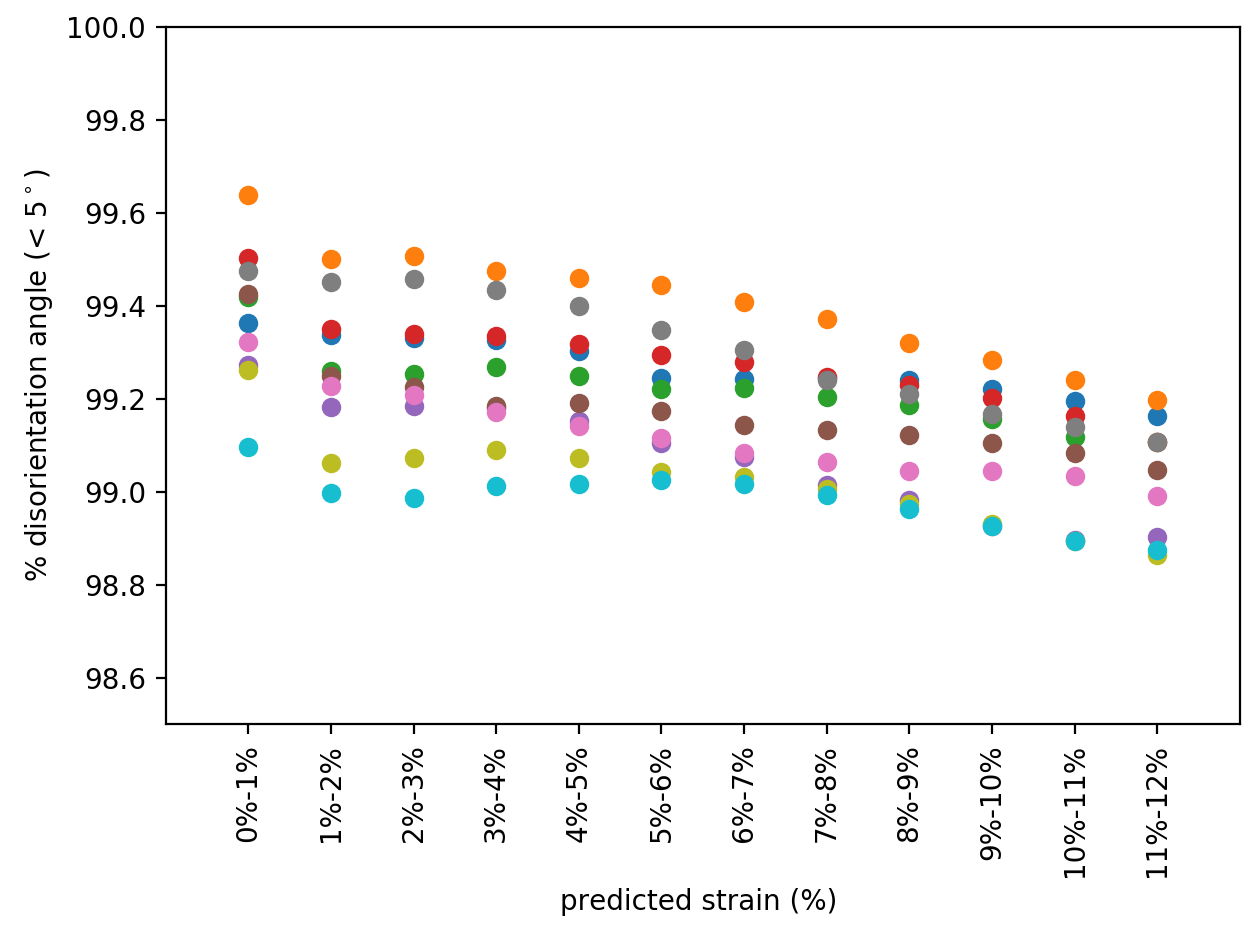}
 }\qquad
\subfloat[LSTM-CP Model II \label{stat_syn646464_mixed}]{%
\includegraphics[scale=.65]{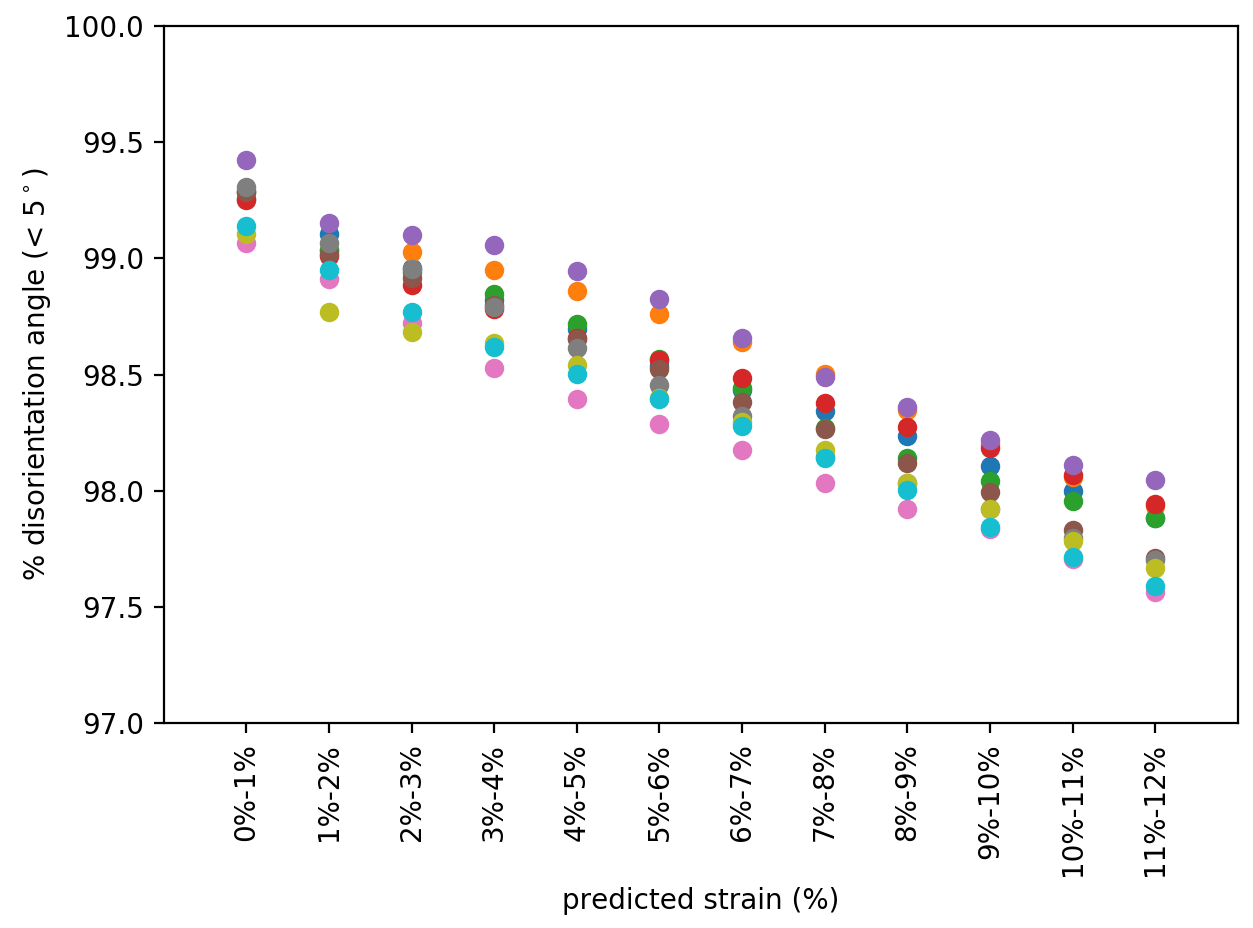}
  }%
\caption{Percentage of disorientation angles less than 5$^{\circ}$ for 10 different 64$\times$64$\times$64 synthetic microstructures (denoted by colors) for the LSTM-CP Model I  and Model II predictions at different strain state. The numbers in X-axis corresponds to the evolution from the current strain state to the next state.}
\label{stat_syn646464_all}
\end{figure}

 As in the case of 16$^3$ 3D synthetic structure, we predicted the 1\% strain evolution, up to 12\% strain, in the initial structure (unstrained or EVPFFT evolved structure at certain strain level) for the 10 random 64$^3$ 3D synthetic structures using both models, which are shown in Fig.\ref{stat_syn646464} and Fig.\ref{stat_syn646464_mixed}, respectively. For Model I and Model II, for all structures and all the strain levels, more than 98\% and 97.5\% of disorientation angles are within 5$^{\circ}$, respectively. Unlike in 16$^3$ structures (as seen in Fig. \ref{stat_syn161616_all}), the model predictions for 64$^3$ for all the strain levels are well within the errors. Summary of these results are tabulated in Table \ref{table_stat_syn161616_646464_all}.
 
 One apparent reason for the increase in the accuracy for  64$^3$  is due to more number of grains and larger grain size in the bigger structure compared to the smaller one. Larger grain size shows more uniform behavior in the local environment compared to the smaller grains. While training the model with 16$^3$ structures, the data sets are large enough to capture the local homogeneity, which is clearly missed for certain strain levels during the reconstruction of the individual microstructures (Fig.\ref{stat_syn161616_all}).
 %

\begin{table}[width=.9\linewidth,cols=4,pos=h]
\caption{Mean and standard devaition of percentage of disorientation angle within 5$^{\circ}$ between the ground truth and  LSTM-CP Model I and Model II predictions at 12 different strain levels for 10 random 64$^{3}$ structures shown in Fig. 10.}\label{tbl5}
\begin{tabular*}{\tblwidth}{@{} LLLLL@{} }
\toprule
          & Model I&&Model II& \\
\midrule
\% Change in strain      &Mean(\%) &Standard deviation  & Mean (\%) & Standard deviation   \\
\midrule
 0\%-1\% &99.38 &0.143 &  99.24 &0.102   \\
  1\%-2\% & 99.26 &0.149  &  99.01 &0.102   \\
  2\%-3\% & 99.26 &0.152  &  98.90 &0.127  \\
  3\%-4\% & 99.25 &0.140  &  98.78 &0.149 \\
  4\%-5\% & 99.23 &0.135  &  98.66 &0.155 \\
  5\%-6\% & 99.20 &0.129  &  98.53 &0.156 \\
  6\%-7\% & 99.18 & 0.124  &  98.41 &0.146 \\
  7\%-8\% & 99.15 &0.122&  98.28 & 0.147  \\
  8\%-9\% & 99.13 &0.122  &  98.15 & 0.144 \\
  9\%-10\% & 99.10 & 0.126  &  98.03 & 0.139 \\
  10\%-11\% & 99.07 &0.124  & 97.90 & 0.145 \\
  11\%-12\% & 99.04 &0.115 & 97.79 &0.156 \\
\bottomrule
\end{tabular*}
\label{table_stat_syn161616_646464_all}
\end{table}
 %

\subsection{LSTM-CP prediction for measured 3D microstructure}
\begin{figure}
\centering
\subfloat[ Disorientation between the EVPFFT simulations of experimental Cu to 1\% strain and LSTM-CP Model I prediction. The mean is 1.03$^\circ$ and the standard deviation of the distribution is 1.203$^\circ$.\label{misorexpfft}]{%
\includegraphics[scale=.40]{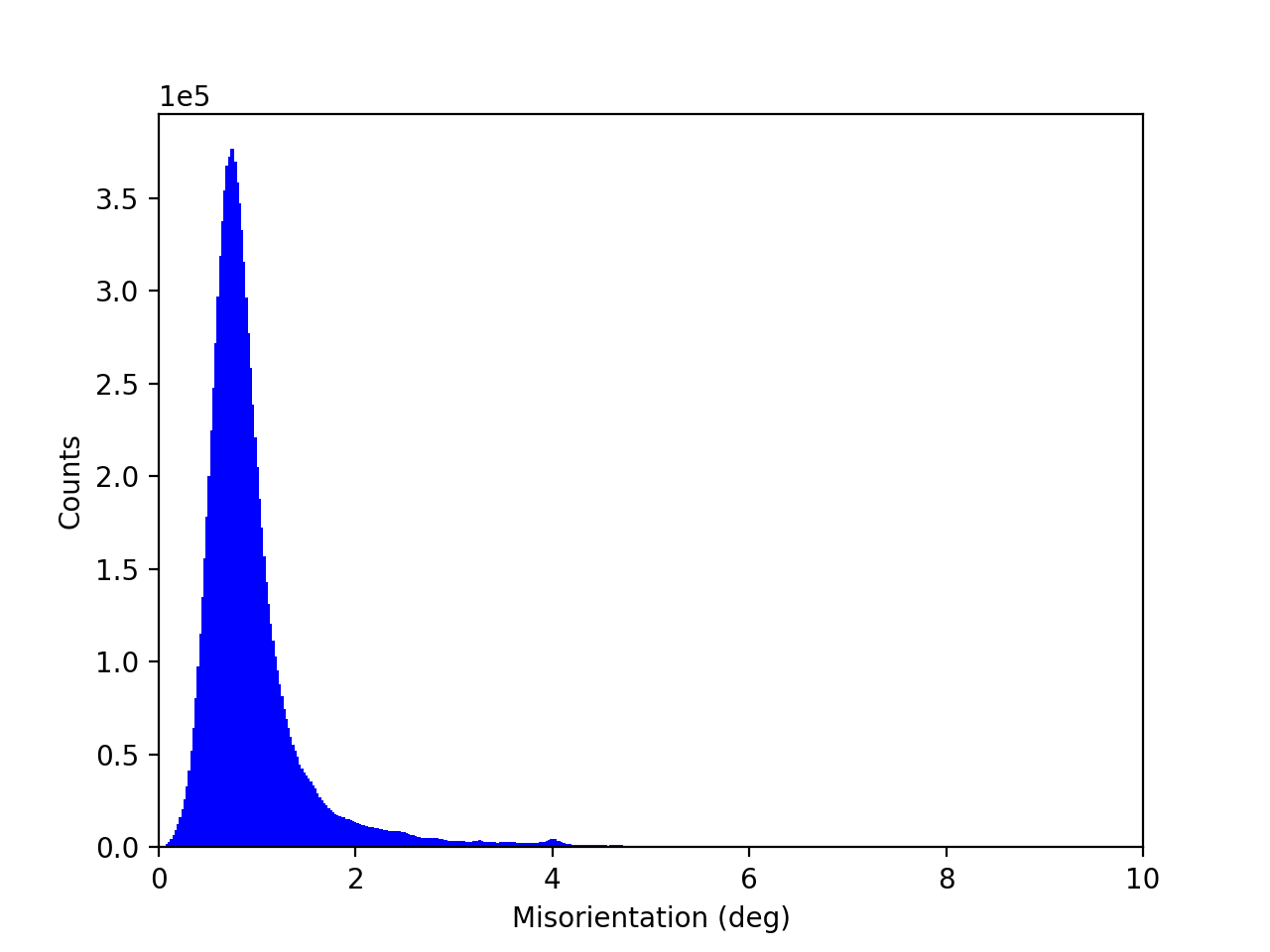}
 }\qquad
 \subfloat[Disorientation between the EVPFFT simulations of experimental Cu to 1\% strain and LSTM-CP Model II prediction. The mean is 1.04 $^{\circ}$ and the standard deviation of the distribution is 0.767$^\circ$. \label{misorexpfft_mixed}]{%
\includegraphics[scale=.40]{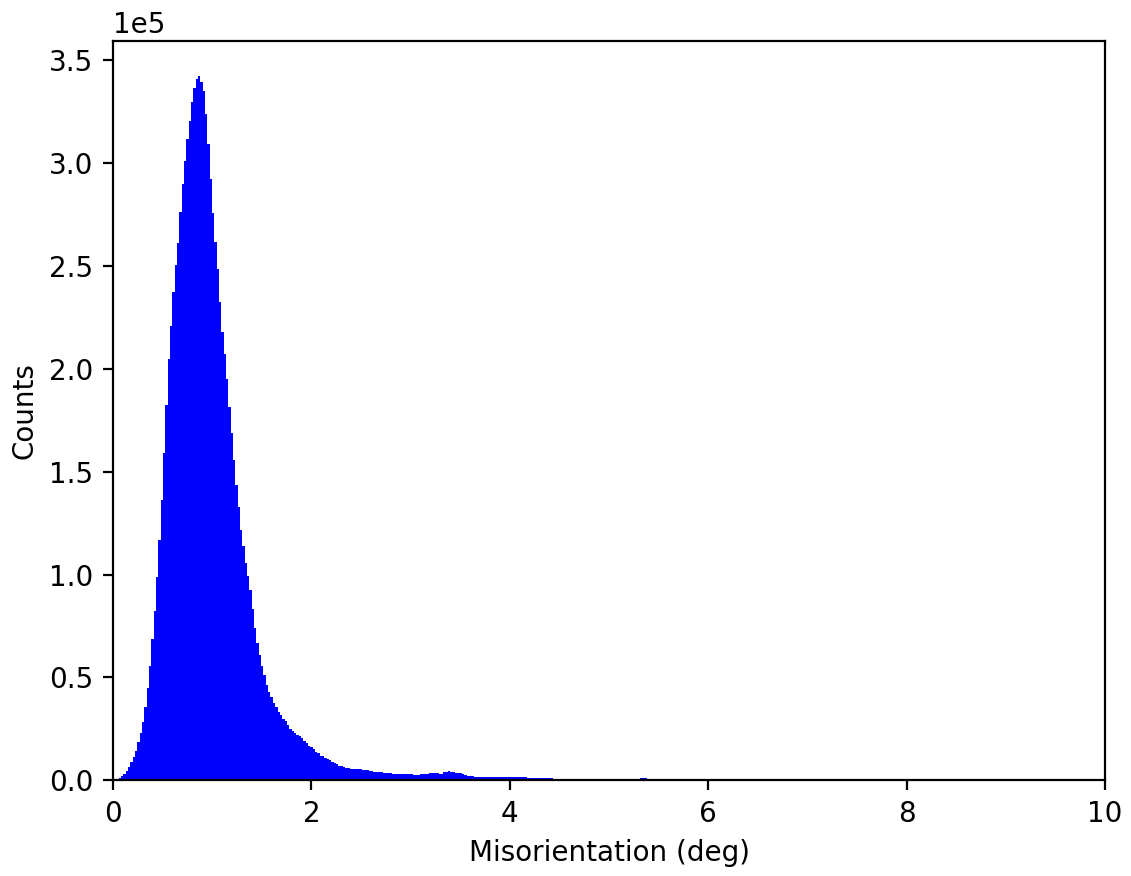}
 }\qquad
 \subfloat[Misorientation map between the predicted and the ground truth (EVPFFT) crystal orientations of experimental Cu for three different layers (layers 25, 50, and 75). The grid size of measured microstructure is 420$\times$420$\times$100. The sample is cylindrical and the buffer region has zero misorientation. \label{misormap_expfft}]{%
\includegraphics[scale=.65]{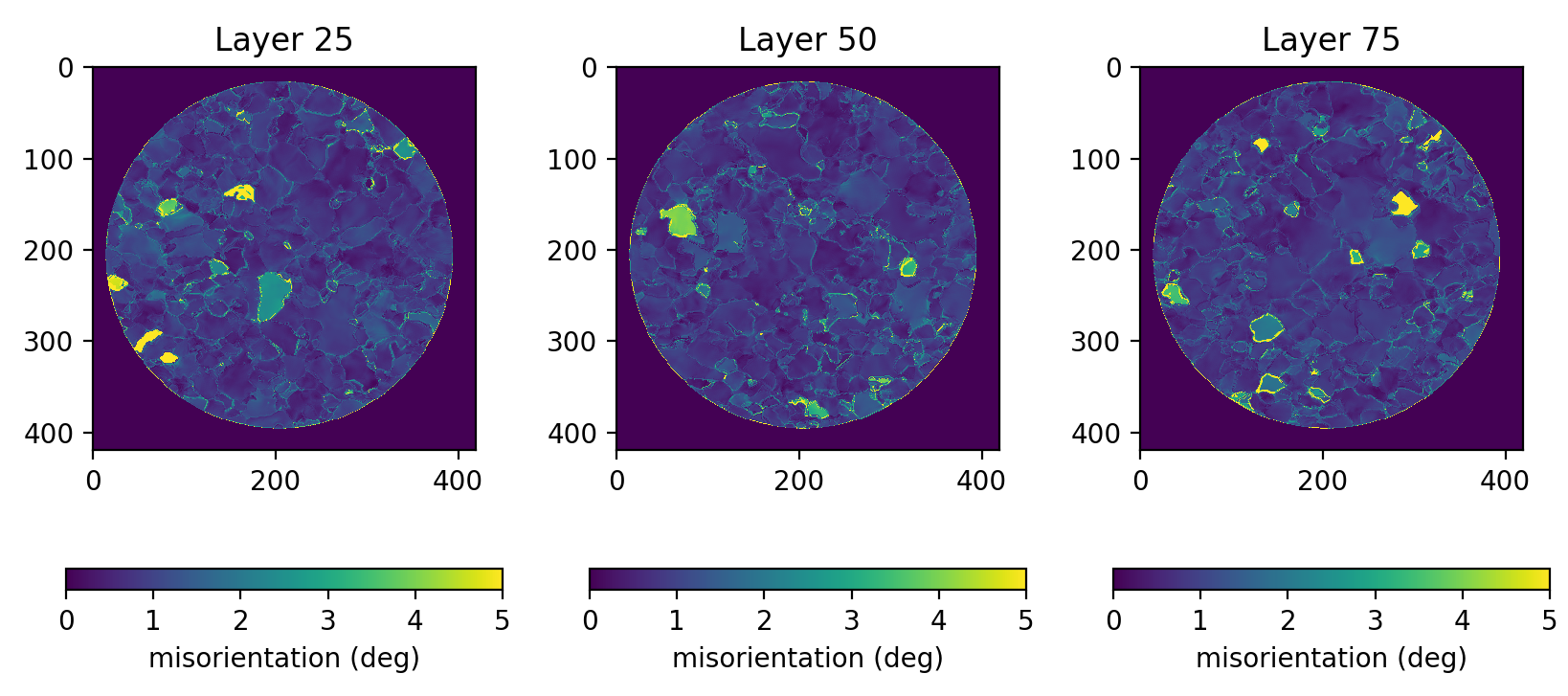}
 }\qquad
\subfloat[ Misorientation map between the LSTM-CP Model II predicted and the ground truth (EVPFFT) crystal orientations of experimental Cu for three different layers (layers 25, 50, and 75). The grid size of measured microstructure is 420$\times$420$\times$100. The sample is cylindrical and the buffer region has zero misorientation.\label{misormap_expfft_mixed}]{%
\includegraphics[scale=.65]{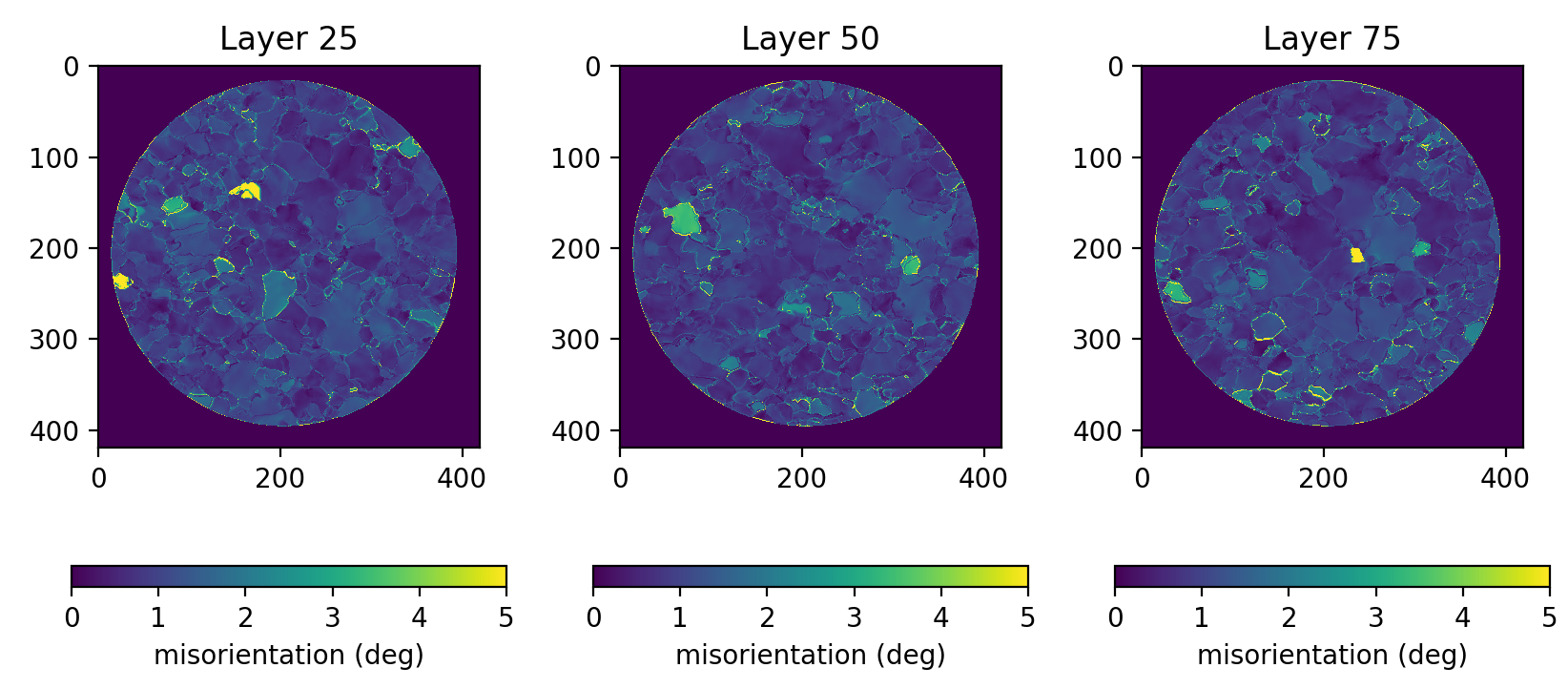}
  }%
\caption{Misorientation distribution for the LSTM-CP Model I and Model II predicted microstructure and the misorientation maps  for the HEDM measured Cu microstructure.}
\label{misorexpfft_all}
\end{figure}

To test the generality of the model, we predicted the 1\% strain evolution of the experimentally measured 3D microstructure obtained from high-energy Xray diffraction microscopy (HEDM) measurements of Cu reported by ~\cite{pokharel2015situ} using  the LSTM-CP Model I and Model II. Experimentally measured microstructures from HEDM was directly used as an input in EVPFFT simulations. As in the case of synthetic structure, EVPFFT simulations of experimental Cu to 1\% strain is used as the `ground truth' for comparison. The disorientation angle distribution between Model I and Model II predictions and the ground truth are shown in Fig. \ref{misorexpfft} and Fig. \ref{misorexpfft_mixed}, respectively. For the measured microstructure, 98.21\% and 99.43\% of the disorientation angles are within 5$^{\circ}$ for Model I and Model II respectively. For Model I, the mean of the distribution is 1.03$^\circ$ and the standard deviation is 1.203$^\circ$, and for Model II, the mean is 1.04 $^{\circ}$ and the standard deviation is 0.767$^\circ$.

\begin{figure}
	\centering
		\includegraphics[scale=.60]{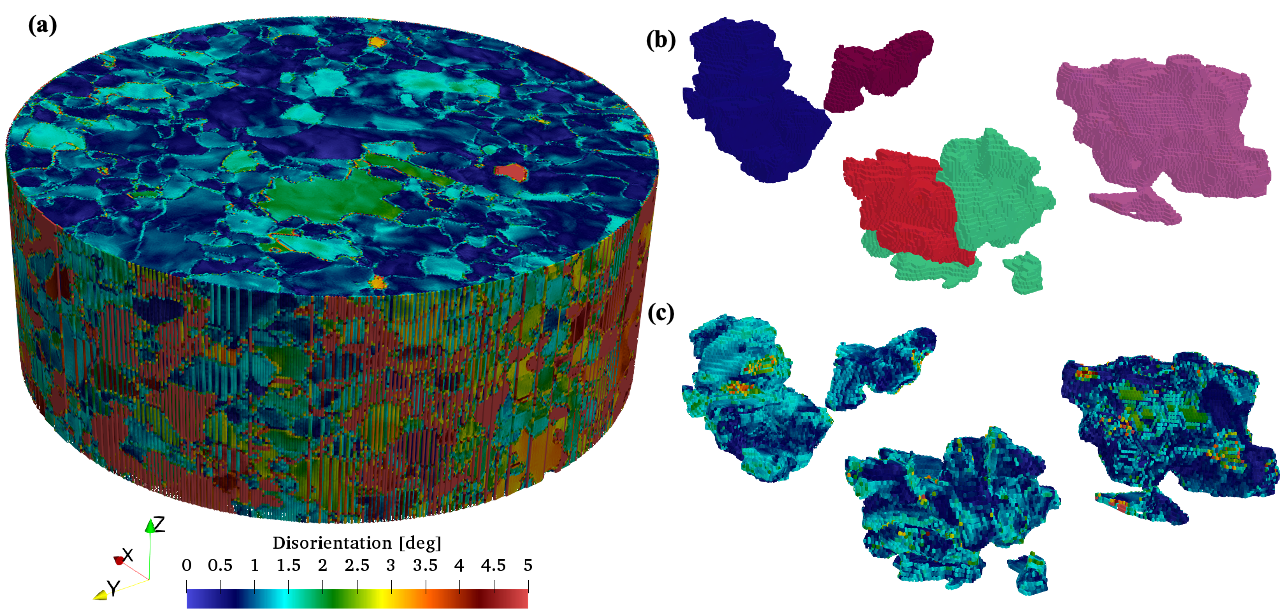}
	\caption{(a) Distribution of disorientation between 1\% EVPFFT simulated experimental Cu microstructure and the LSTM-CP Model II. Color scheme shows the disorientation angle in degrees. (b) Orientations of random grains. (c)Disorientation  distribution in random grains shown in (b).}
	\label{grain_expCu}
\end{figure}

The 2D misorientation maps for three different layers (layers 25, 50, and 75) are shown in Fig. \ref{misormap_expfft} and Fig. \ref{misormap_expfft_mixed}. The experimental Cu sample is cylindrical, and the misorientation angle of zero is assigned to the buffer region. The misorientation maps confirms the low disorientation angles distribution for both the models. The disorientation map for the full 3D microstructure is shown in Fig. \ref{grain_expCu}(a). The accuracy of the predictions at the grain scale are also evaluated for five randomly selected grains from the 3D microstructure. Fig. \ref{grain_expCu}(b) shows the grains colored by crystal orientation and Fig. \ref{grain_expCu}(c) shows the disorientation map. All 3D grains show low disorientation angle distribution.
%
\begin{table}[width=.9\linewidth,cols=4,pos=h]
\caption{Mean and standard deviation of disorientation angle between the ground truth and  LSTM-CP Model I and Model II predictions of 16$^{3}$, 64$^{3}$ and experimental Cu structures.}\label{tbl_mean_sd}
\begin{tabular*}{\tblwidth}{@{} LLLLL@{} }
\toprule
          & Model I&&Model II& \\
\midrule
Structure      &Mean($^{\circ}$) &Standard deviation  & Mean ($^{\circ}$) & Standard deviation   \\
\midrule
 16$\times$16$\times$16&1.31 &0.821 &  0.96 &0.601  \\
 64$\times$64$\times$64 & 1.04 &0.916  &  1.12 &0.852   \\
 Exp. Cu & 1.03 & 1.203  &  1.04 &0.767  \\
\bottomrule
\end{tabular*}
\end{table}
 %

\subsection{Parametric study of the strain step size}
\begin{figure}
\centering
\subfloat[The LSTM-CP Model III predicted Euler angles versus the ground truth (EVPFFT) trained for the sequential data of 2\% strain from 16$\times$16$\times$16 synthetic microstructures.\label{lstm_pred_m3}]{%
\includegraphics[scale=.60]{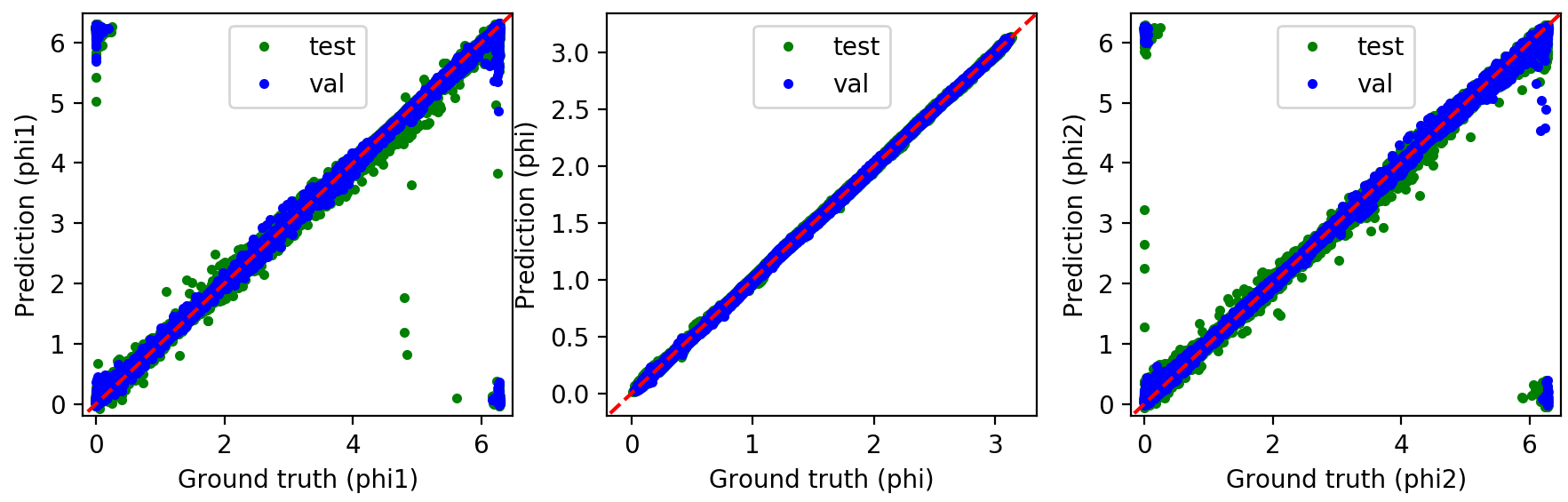}
 }\qquad
\subfloat[The LSTM-CP Model IV predicted Euler angles versus the ground truth (EVPFFT) trained for the sequential data of 3\% strain from 16$\times$16$\times$16 synthetic microstructures.\label{lstm_pred_m4}]{%
\includegraphics[scale=.60]{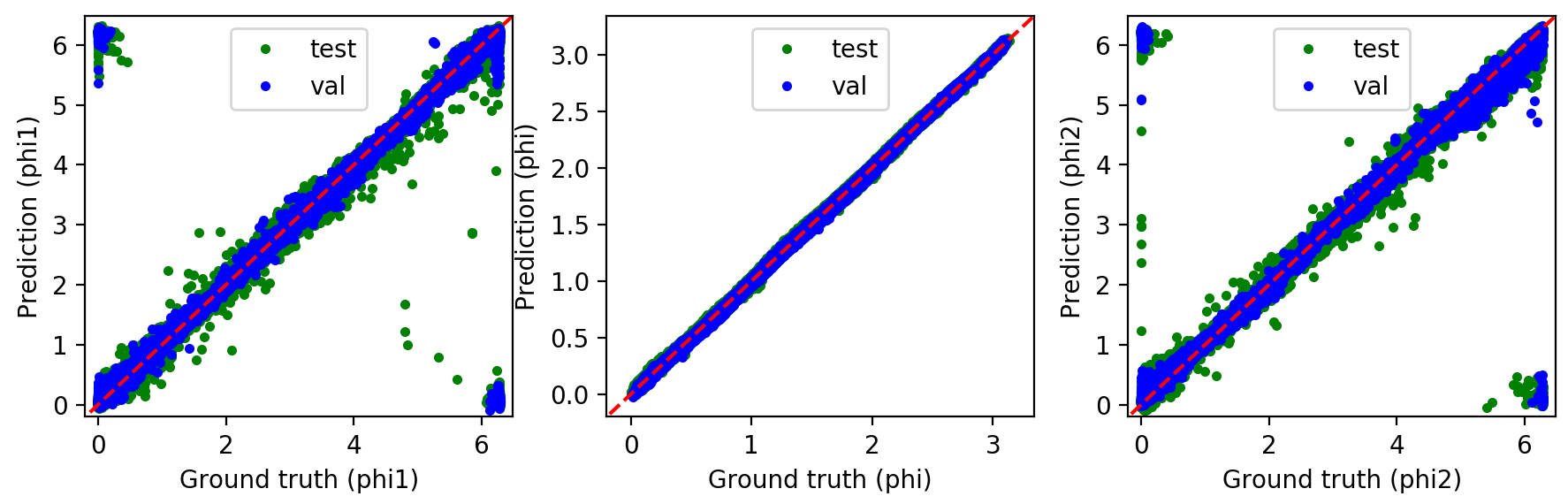}
  }%
\caption{The LSTM-CP predicted Euler angles versus the ground truth (EVPFFT) for Model III and Model IV. }
\label{lstm_pred_param}
\end{figure}

\begin{figure}
\centering
\subfloat[Comparison of disorientation distributions from three different models predictions for a 16$^{3}$ synthetic structure. 99.38\%, 98.32\% and 52.32\% of  disorientation angles are within 5$^{\circ}$ for Model I, Model III and Model IV, respectively. The mean and standard deviation of the distribution for Model I are 1.31$^{\circ}$ and 0.821$^{\circ}$, respectively. The mean and standard deviation of the distribution for Model III are 2.38$^{\circ}$ and 1.01$^{\circ}$, respectively. The mean and standard deviation of the distribution for Model IV are 5.15$^{\circ}$ and 2.41$^{\circ}$, respectively. \label{misor_diff_models_syn}]{%
\includegraphics[scale=.50]{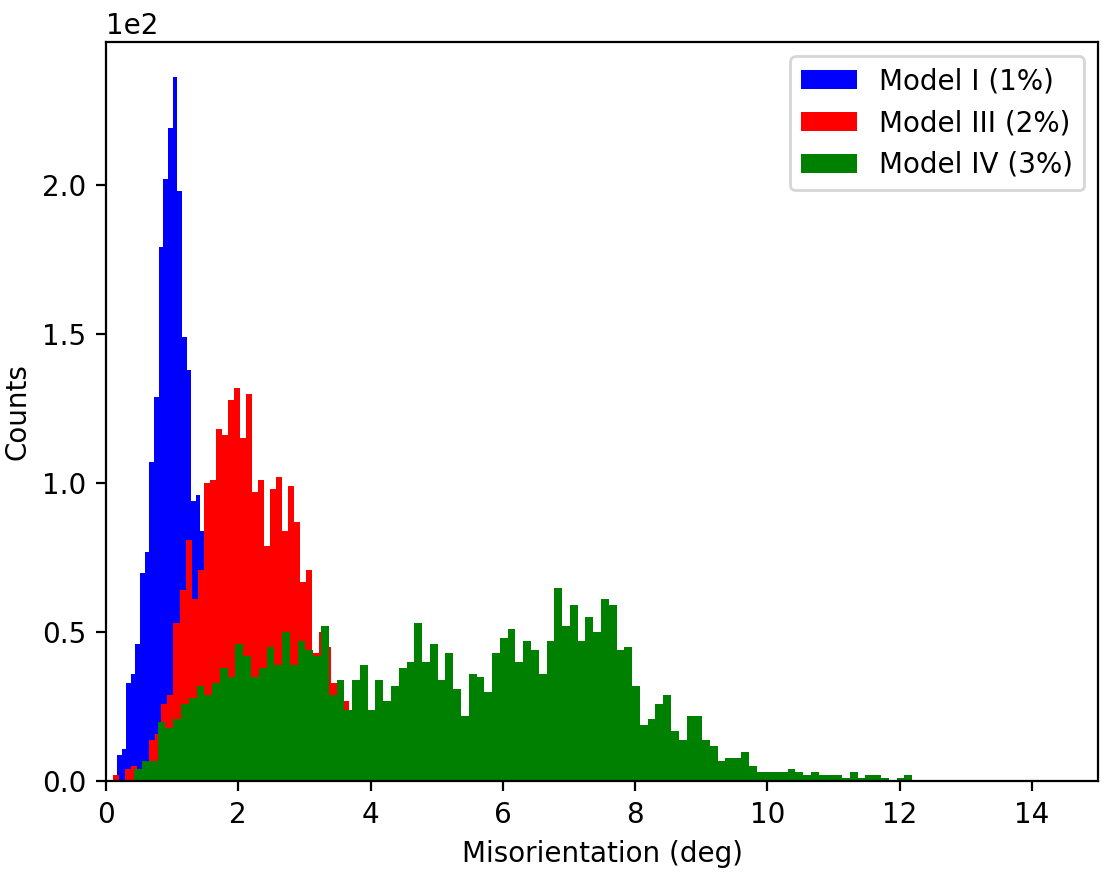}
 }\qquad
\subfloat[Comparison of disorientation distributions from three different models predictions for an experimental Cu microstructure.  98.21\%, 97.08\% and 94.42\% of  disorientation angles are within 5$^{\circ}$ for Model I, Model III and Model IV,  respectively. The mean and standard deviation of the distribution for Model I are 1.03$^{\circ}$ and 1.203$^{\circ}$, respectively. The mean and standard deviation of the distribution for Model III are 1.77$^{\circ}$ and 1.240$^{\circ}$, respectively. The mean and standard deviation of the distribution for Model IV are 2.72$^{\circ}$ and 1.998$^{\circ}$, respectively.\label{misor_diff_models_exp}]{%
\includegraphics[scale=.50]{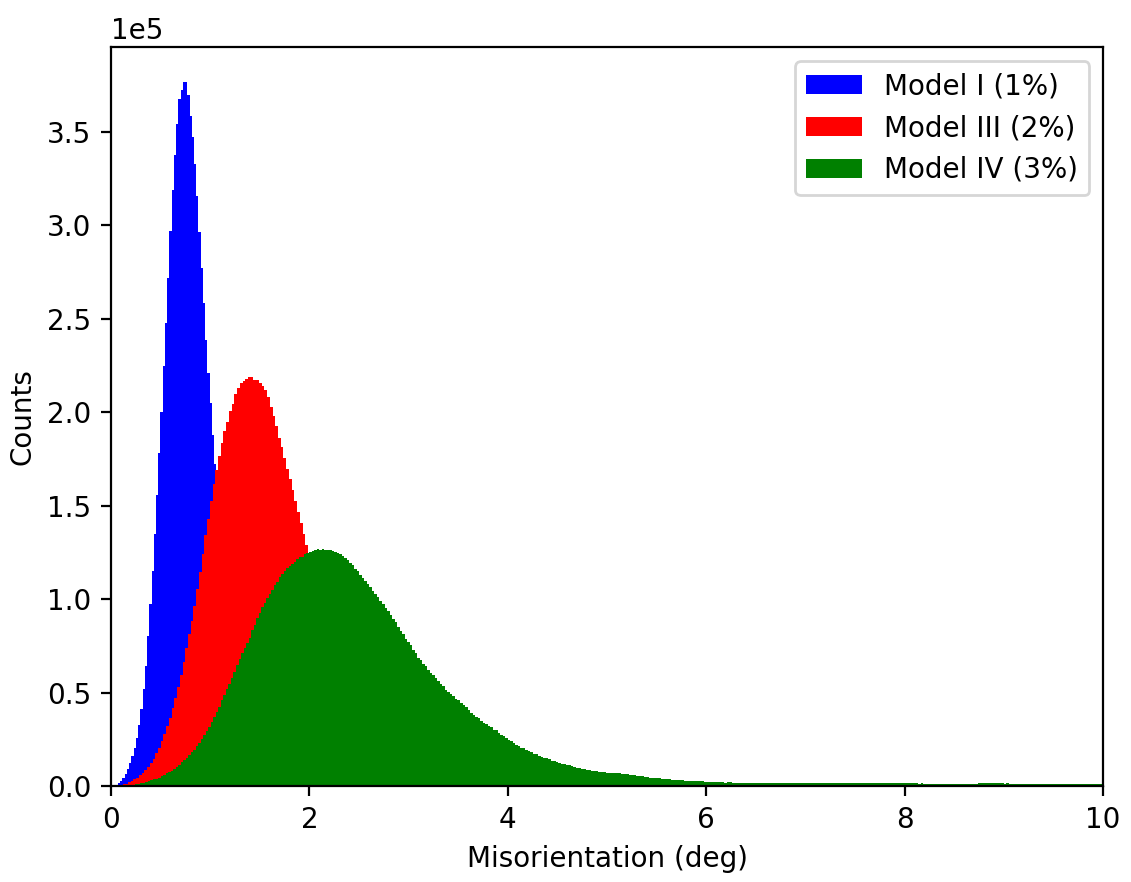}
  }%
\caption{Comparison of disorientation angles distributions from three different models for the synthetic structure and the experimental Cu microstructure. Model I is for 1\% , Model III is for 2\% and Model IV is for 3\% strain predictions.}
\label{misor_diff_models_all}
\end{figure}

Model III and Model IV predictions for the test set and the validation set at the end of the training are shown in Fig. \ref{lstm_pred_param}. It can be seen that the accuracy of the model decreases with the increase in strain step. Fig. \ref{misor_diff_models_all}  shows the comparison of disorientation distributions for the three models (Model I, Model III and Model IV) for the synthetic structure (Fig. \ref{misor_diff_models_syn}) and for the experimentally measured Cu microstructure (Fig. \ref{misor_diff_models_exp}). For both cases, the distribution gets broader as the strain step increases. For the synthetic structure, 99.38\%, 98.32\% and 52.32\% of the disorientation angles are within 5$^{\circ}$ for Model I, Model III, and Model IV, respectively. The mean and standard deviation of the distributions for Model I are 1.31$^{\circ}$ and 0.821$^{\circ}$, for Model III are 2.38$^{\circ}$ and 1.01$^{\circ}$, and for Model IV are 5.15$^{\circ}$ and 2.41$^{\circ}$. Thus, Model IV fails in accurately predicting the microstructure evolution. 

The summary of the results are shown in Table \ref{tbl4}. For the experimental Cu microstructure, 98.21\%, 97.08\% and 94.42\% of disorientation angles are within 5$^{\circ}$ for Model I, Model III and Model IV, respectively.  The mean and standard deviation of the distributions for Model I are 1.03$^{\circ}$ and 1.203$^{\circ}$, for Model III are 1.77$^{\circ}$ and 1.240$^{\circ}$, and for Model IV are 2.72$^{\circ}$ and 1.998$^{\circ}$. The experimental Cu and 64$^3$ synthetic structure has higher percentage of disorientation angles within 5$^{\circ}$ for Model III and Model IV compared to the 16$^3$ synthetic structures. This could be due to statistical effect of having larger grains with many interior voxels that are similar in the experimental structure and the 64$^3$ structure.

\begin{table}[width=.9\linewidth,cols=4,pos=h]
\caption{Percentage of disorientation angle within 5$^{\circ}$ between the ground truth and three different LSTM-CP model predictions .}\label{tbl4}
\begin{tabular*}{\tblwidth}{@{} LLLL@{} }
\toprule
          & &Disorientation angles (<5$^{\circ}$) & \\
\midrule
      &Synthetic structure  & Synthetic structure & Experimental Cu   \\
      & (16$\times$16$\times$16) & (64$\times$64$\times$64)&(420$\times$420$\times$100)\\
\midrule
 Model I (1\% strain) & 99.38 & 99.43  & 99.04  \\
 Model III (2\% strain) & 98.32 & 97.58 & 98.09 \\
 Model IV (3\% strain) & 52.32 & 93.28  & 94.88 \\
\bottomrule
\end{tabular*}
\end{table}
%

%
%
%
%

\section{Discussion}

We have presented a method that captures local heterogeneity and non-linearity inherent in polycrystalline materials and predicts microstructure evolution in 3D, without having to perform expensive CP simulations. Our working hypothesis and the motivation for the approach taken in this work is: in a physical system, whose dynamics are locally governed by forces that are the gradients of potential fields, only local interactions are needed to train a surrogate model. As a first-order approximation, we consider only a local effect on each crystallite point by taking only the nearest neighbors (up to the 3rd nearest neighbors are accounted for in the training data) interactions. Any long-range interactions are not explicitly accounted for in this proof-of-concept work. Adopting this hypothesis, predicting the 3D microstructure evolution problem became tractable by significantly reducing the amount of training data required for building a model that can predict the spatial evolution of microstructural parameters in 3D. Furthermore, once our model is trained, it is independent of the test microstructure to which it can be applied because all of the calculations are local in nature. This was demonstrated on an experimentally measured Cu microstructure, which has different grain size, grain morphology, texture, and internal stresses in comparison to the synthetic structures, and yet the LSTM-CP model that was trained only on synthetic structures was successful in predicting the evolution of orientation under a uniaxial tensile deformation up to the level of CP. This illustrates the robustness of this LSTM-CP modelling approach in predicting the 3D spatial orientation evolution under uniaxial tensile strain for a variety of microstructures. This has otherwise been a major limitation in the research reported in the current literature. 

The microstructure evolution predicted from Model I and Model II are almost identical for all the synthetic structures, as well as the experimental Cu structure. This signifies the relevance of incorporating only local interactions during the training process. Small representative structures with a grid size $16^3$ have fewer grains and yet the local environment on each crystallite resembles the real system and is sufficient to capture the complex phenomenon of 3D microstructure evolution. For the study of orientation evolution under uniaxial strain, models trained from 16$^{3}$ structures (Model I) were found to provide accurate predictions. The parametric study on strain step size shows that there should be a cut-off in step size at approximately 2\% strain, information embedded in the data is lost when the step size is much larger. However, this cut-off step size for the LSTM-CP model is still significantly larger than what is possible with the small-strain EVPFFT model used in this work. 

\subsection{Computational efficiency}
There have been many studies aimed at lowering computational demand to accelerate full-field crystal plasticity simulations~\cite{savage2015computer,eghtesad2020multi}. Because our trained model can instantly evolve a microstructure, it eliminates the need for large computing resources which are required for the iterative methods of conventional CP calculations, which must take large numbers of very small strain steps ($\sim$0.02\%) in order to ensure convergence. 

Both LSTM-CP and EVPFFT can be run in parallel on thousands of cores in high performance computing (HPC) clusters. To compare the speed of LSTM-CP relative to EVPFFT we ran both on a single 2.5 GHz Intel Xenon W processor. The LSTM-CP model showed a speed up of $>6\times$ in predicting polycrystalline microstructure evolution under plastic deformation in comparison with EVPFFT for a single 0.02\% strain step and a speed up of  $>312\times$ for 1\% strain evolution. 
A maximum strain step size of EVPFFT is $\sim$0.02\%, taking steps bigger than this causes the code to not converge and the predictions become inaccurate. For a 420$\times$420$\times$100 voxel experimental Cu microstructure EVPFFT requires approximately 40 minutes of computational time for a single 0.02\% strain step. In comparison, a single step of the LSTM-CP model on the same computer and with the same microstructure requires 6.4 minutes. If we are interested in observing very high resolution of strain steps (0.02\%) evolution of a microstructure, the LSTM-CP method provides an increase of at least a factor of 6. 

Our studies have shown that the LSTM-CP model can take single steps as large as 2\% before the prediction accuracy starts to degrade. In practice, individual 3D measurements at beam times, such as the HEDM method, are extremely time consuming (hours) and therefore experiments observe structural evolution at much lower resolution in terms of strain steps, typically steps of $\sim$1\% are taken. Therefore, if we are interested in observing lower resolution strain steps of $\sim$1\%, for comparison to experimental data, the EVPFFT approach requires $50\times0.02$\% steps to simulate a 1\% evolution and this would require $50\times40$ minutes $\approx$ 33.3 hours on the single CPU setup described above. For the same 1\% evolution the LSTM-CP approach would still only require 6.4 minutes for a single 1\% steps, which in this case is a speed up of $>$300$\times$.

For taking large strain steps LSTM-CP models avoid the use of high performance computer clusters in crystal plasticity simulations, making them applicable for use in real-time in parallel with running experiments using only a single local desktop machine. For higher resolution strain step studies, LSTM-CP can be used on an HPC cluster maintaining a $>6\times$ speed up compared to EVPFFT.

\subsection{Higher accuracy hybrid approach}
Although the LSTM-CP approach is orders of magnitude faster than the traditional iterative CP method, it is not quite as accurate, as seen from the results above. However, a hybrid approach combining LSTM-CP with traditional iterative CP methods has the potential to provide extremely large speed ups, as described above, without compromising accuracy relative to a traditional CP method alone. For example, in order to quickly calculate the evolution of a sample state $S_0$ through a 2\% strain step to state $S_1$, we can use the LSTM-CP model to almost instantly provide an accurate estimate $\hat{S}_1$ of $S_1$ and then use that as a starting point for a traditional iterative CP calculation, which will fine tune $\hat{S}_1$ to create $S_1$ in a much smaller number of steps than would have been required if it started from $S_0$. 

\subsection{Limitations and future work}
The current study was limited to orientation evolution under uniaxial strain, and local interactions were shown to be sufficient to describe such a phenomenon accurately. In future studies, we aim to develop a model that can predict mechanical properties such as stress and strain fields~\cite{ali2019application} together with the local microstructure information. The current models can predict an evolution for 1\% strain from any given strain level up to 12\% strain from a simple vanilla LSTM. In future work, the idea of locality, in conjunction with multiple time series LSTM will be explored for time series studies of microstructure evolution. 

Furthermore, the LSTM-CP model provides the flexibility of incorporating experimental data to improve its predictive accuracy by re-training and transfer learning, as recently demonstrated in \cite{shen2019convolutional}. In \cite{shen2019convolutional} the authors applied a convolutional neural (CNN) network to map electron back scatter diffraction (EBSD) patterns to quaternions representing crystal orientations. Their CNN was first trained using large quantities (hundreds of thousands) of readily generated simulation data. The model was then extended for use with experimental data by utilizing two techniques: transfer learning and re-training. Such an approach can be taken here as well to improve the accuracy of the LSTM-CP model's predictions for new experimentally collected data sets. One goal of our future studies is to collect many more 3D HEDM-based measurements at a high resolution of strain steps ($<1\%$) and to use that data to further improve the predictive power of our models.

In this work, disorientation angles were considered as the metric to test model prediction accuracy relative to CP predictions. In the data presented above, we reported the percent of voxels whose orientations were predicted to be within 5$^{\circ}$ of CP predictions and showed that for both 1\% and 2\% strain steps more than 97\% of our prediction errors were <5$^\circ$ with an average close to $\sim$1$^\circ$. The reason for the choice of 5$^\circ$ is that this is the currently achievable accuracy of CP models when applied to predict actual experimentally measured volume orientation evolution at the local sub-grain scale level \cite{pokharel2014polycrystal}.

We also note that in the present proof-of-concept work, the elastic stiffness parameters were chosen for a single crystal Cu and Voce hardening parameters for EVPFFT simulations were obtained from calibrating the stress strain curves reported in ~\cite{pokharel2015situ}. Therefore, each model will have to be trained with material-specific parameters for accurate predictions, to the level of EVPFFT.

\section{Conclusions}
We have applied long short-term memory, an artificial recurrent neural network, to predict the microstructure evolution of polycrystalline metals. The predictions are compared to the crystal plasticity fast Fourier transform-based model. The model provided a significant gain in the computational time with the same accuracy as CPFFT. We have successfully demonstrated that the model trained from a representative synthetic data can predict the deformation on the 3D experimental data from HEDM to the level of accuracy of CPFFT. We have shown how the model easily transfers to the simulations of experimental HEDM data. The main novel contributions of this work can be summarized as:

\begin{enumerate}
	\item	The ML-based surrogate model trained and tested on synthetic 3D microstructure data was as accurate as of the conventional micromechanical model and orders of magnitude faster. Our method does not need to perform numerical iterations per strain step otherwise needed by the conventional method for convergence.

	\item The main strength of this model, apart from significant reduction in computational cost, is that it is completely general and can be used for arbitrary 3D polycrystalline microstructure. This was achieved by implementing a novel method, which accounts for each point's interactions with only its local neighborhood in the training data.

	\item	The LSTM model employed in this work is able to predict the spatially and temporally resolved 3D microstructure evolution under uniaxial tensile loading. 
	
	\item	We demonstrate the applicability of this method on an experimentally measured 3D microstructure of FCC Cu undergoing tensile deformation. A comparison between ML-model predictions and CPFFT simulations shows excellent agreement.
\end{enumerate}

This proof-of-concept work demonstrated that given a previous state microstructure, the surrogate model could accurately predict the 3D grain reorientation due to plastic deformation in the next strain state during tensile loading. Also, the surrogate models developed here were limited to a fixed strain step and a fixed boundary condition. This framework can be extended to predict the evolution of a series of microstructure and micromechanical properties such as strain and stress fields. Furthermore, by generating training sets for various macroscopic conditions, microstructure evolution for a more general loading condition can also be predicted.

\section*{Acknowledgements}
This work was supported by the U.S. Department of Energy through the Los Alamos National Laboratory. Los Alamos National Laboratory is operated by Triad National Security, LLC, for the National Nuclear Security Administration of U.S. Department of Energy (Contract No. 89233218CNA000001). This work was supported by Los Alamos National Laboratory Laboratory Directed Research and Development (LDRD) project $\#$20190571ECR. We are grateful to Dr. A. Scheinker and Dr. A.D. Rollett for fruitful discussions.

\bibliographystyle{cas-model2-names}

\bibliography{cas-refs}

\end{document}